\documentclass[twocolumn]{aastex631}

\usepackage{comment}
\usepackage{graphicx}
\usepackage{txfonts}
\newcommand{\rev}[1]{{\textcolor{black}{#1}}}
\newcommand{\eqref}[1]{(\ref{#1})}

\shorttitle{Tracing Nitrogen In Giant Planet Atmospheres I}
\graphicspath{{./}{figures/}}
\begin{document} 

%%%%%%%%%%%%%%%%%%% TITLE PAGE %%%%%%%%%%%%%%%%%%%
%\title{Universal Deep Adiabat in Warm Exoplanetary Atmospheres: Implications for Nitrogen Disequilibrium Chemistry and Bulk Nitrogen Tracers in Spectroscopic Observations}
%\title{Trace Nitrogen: Insights from Nitrogen Disequilibrium Chemistry and Universal Deep Adiabat in Warm Exoplanetary Atmospheres}
%\title{Nitrogen as a New Dimension of Planet Formation Tracer: Can We Diagnose Bulk Nitrogen from Observations of Ammonia in Exoplanetary Atmospheres?}
\title{Nitrogen as a Tracer of Giant Planet Formation. I.: A Universal Deep Adiabatic Profile and Semi-analytical Predictions of Disequilibrium Ammonia Abundances in Warm Exoplanetary Atmospheres}
%\title{Nitrogen as a New Dimension of Planet Formation Tracer: Strategy to Constrain Bulk Nitrogen Abundance of Warm Exoplanetary Atmospheres}
%\title{Beyond C/O Ratio: How Observable NH$_3$ Relates with Bulk Nitrogen Abundance of Exoplanetary Atmospheres}

%\title{Can Ammonia be Used as a Bulk Nitrogen Tracer? Insights from Disequilibrium Chemistry in a Universal Deep Adiabat of Warm Exoplanetary Atmospheres}

%\title{Can We Use Ammonia as a Bulk Nitrogen Tracer in Giant Planets?}

\shortauthors{Ohno \& Fortney}

\author[0000-0003-3290-6758]{Kazumasa Ohno}
\affiliation{Department of Astronomy \& Astrophysics, University of California, Santa Cruz, 1156 High St, Santa Cruz, CA 95064, USA}

\author[0000-0002-9843-4354]{Jonathan J. Fortney}
\affiliation{Department of Astronomy \& Astrophysics, University of California, Santa Cruz, 1156 High St, Santa Cruz, CA 95064, USA}

%\author{Neel Patel}
%\affiliation{Department of Astronomy \& Astrophysics, University of California, Santa Cruz, 1156 High St, Santa Cruz, CA 95064, USA}
%%%%%%% Abstract %%%%%%%%%%%%%%
\begin{abstract}
%Atmospheric compositions provide valuable clues to the planetary formation and evolution processes.
\rev{A major motivation of spectroscopic observations of giant exoplanets is to unveil planet formation processes from atmospheric compositions.
Several recent studies suggested that atmospheric nitrogen, like carbon and oxygen, can provide important constrains on planetary formation environments.
Since nitrogen chemistry can be far from thermochemical equilibrium in warm atmospheres, 
%and the N$_2$ molecule has negligible opacity at infrared wavelengths for the temperature regime of exoplanets, 
we extensively investigate under what conditions, and with what assumptions, the observable NH$_3$ abundances can diagnose an atmosphere's bulk nitrogen abundance. 
In the first paper of this series, we investigate atmospheric T-P profiles across equilibrium temperature, surface gravity, intrinsic temperature, atmospheric metallicity, and C/O ratio using a 1D radiative-convective equilibrium model.
Models with the same intrinsic temperature and surface gravity coincide with a shared ``universal" adiabat in the deep atmosphere, across a wide equilibrium temperature range (250-1200 K), which is not seen in hotter or cooler models.  We explain this behavior in terms of the classic ``radiative zero solution" and then
establish a semi-analytical T-P profile of the deep atmospheres of warm exoplanets.  This profile is then used to predict vertically quenched NH$_3$ abundances.
At solar metallicity, our results show that the quenched NH$_3$ abundance only coincides with the bulk nitrogen abundance (within 10\%) at low intrinsic temperature, corresponding to a planet with a sub-Jupiter mass ($\la1~{\rm M_{\rm J}}$) and old age ($\ga 1~{\rm Gyr}$).  If a planet has a high metallicity ($\ga 10\times$ solar) atmosphere, the quenched NH$_3$ abundance signficantly underestimates the bulk nitrogen abundance at almost all planetary masses and ages.  We suggest modeling and observational strategies to improve the assessment of bulk nitrogen from NH$_3$.}

%Both semi-analytical and numerical calculations demonstrate that, at solar metallicity, the quenched NH$_3$ abundances only coincide with bulk nitrogen abundance for sub-Jupiter mass ($\la1M_{\rm J}$) planets at older ages ($\ga1~{\rm Gyr}$) for a wide range of planetary equilibrium temperature ($\la 800~{\rm K}$) and eddy diffusion coefficients.  The problem is even more pronounced at higher metallicities, which favors N$_2$ over NH$_3$.
%However, to mitigate this problem we construct a semi-analytical prescription to diagnose the bulk nitrogen abundance from the vertically quenched NH$_3$ abundance.
%Furthermore, we perform photochemical calculations for various planetary properties, such as mass, age, and equilibrium temperature. The photodissociation of NH$_3$ potentially depletes observable NH$_3$ abundances significantly.%, though the UV-shielding by photochemical hazes may mitigate the depletion.
%We suggest that warm planets around K- and M-type stars are the best targets for constraining bulk nitrogen abundances through NH$_3$ observations thanks to inefficient NH$_3$ photodissociation. 
%We also compute synthetic transmission spectra of warm Jupiters and predict that NH$_3$ leaves detectable ($>50~{\rm ppm}$) signatures at $1.5$ and $11~{\rm {\mu}m}$ in the equilibrium temperature range of $400$--$1000~{\rm K}$, which are accessible with JWST.
%Our semi-analytical prescription still provides a reasonably guess for 

\end{abstract}
%\keywords{exoplanetary atmosphere ---
%        atmospheric retrieval
%       }
%%%%%%%%%%%%%%%%%%%%%%%%%%%%%%%%%
\section{Introduction}
Planetary atmospheric compositions offer valuable clues to the planet formation and evolution process, \rev{especially for giant planets with primordial atmospheres}.
Over the past decade a number of studies have suggested that atmospheric elemental ratios, such as the carbon-to-oxygen ratio (C/O), can diagnose the orbital distance where a planet initially forms \citep[e.g.,][]{Oberg+11,Madhusudhan+14,Madhusudhan+17,Ali-dib+14,Helling+14,Thiabaud+15,Piso+15,Piso+16,Oberg&Bergin16,Cridland+16,Cridland+17,Cridland+19,Espinoza+17,Eistrup+16,Eistrup+18,Eistrup+22,Booth+17,Booth&Ilee19,Oberg&Wordsworth19,Ohno&Ueda21,Turrini+22,Schneider&Bitsch21,Molliere+22,Pacetti+22,Bitsch+22,Notsu+22,Eistrup22}.
Many previous studies focused on the atmospheric C/O ratio, as it has significant impacts on atmospheric chemistry and likely leaves observable fingerprints \citep[e.g.,][]{Madhusudhan+12,Moses+13,Moses+13b,Molliere+15,Drummond+19,Notsu+20,Dash+22}.
Beyond the C/O ratio, several recent studies also have also discussed the potential importance of other elements, such as nitrogen \citep{Piso+16,Cridland+20,Ohno&Ueda21,Notsu+22}, sulfur \citep{Turrini+22,Pacetti+22}, and refractory metals \citep{Lothringer+21,Schneider&Bitsch21b,Hands&Helled22,Chachan+22}.

Nitrogen is the third most abundant volatile element in solar composition and may provide important constrains on the planetary formation environments.
Nitrogen has particular advantages to probe the formation locations.
\rev{
\citet{Piso+16} first pointed out that the N/O ratio of disk gas is always higher than stellar N/O by a factor of $\ge$2 and monotonically increases with radial distance, which provides additional clues to constrain planetary formation location from a planet's atmospheric N/O ratio.
\citet{Cridland+20} studied the atmospheric compositions of warm Jupiters using a population synthesis model and suggested that combining C/O and N/O helps to probe the formation history, such as whether the planet acquired its atmosphere outside of the refractory carbon erosion front.
\citet{Ohno&Ueda21} also stressed that the atmospheric N/O is expected to be sensitive to formation location if disk solids, such as pebbles and planetesimals, determine the atmospheric composition. 
This is because the solid N/O ratio has an order-of-magnitude variation as a function of a radial distance (see also \citealt{Notsu+22} for the discussion based on a disk chemistry model). 
\citet{Turrini+22} and \citet{Pacetti+22} suggested that C/N, N/O, and S/N ratios help to constrain the formation and migration pathways of giant planets.
}

\begin{figure}[t]
\centering
\includegraphics[clip, width=\hsize]{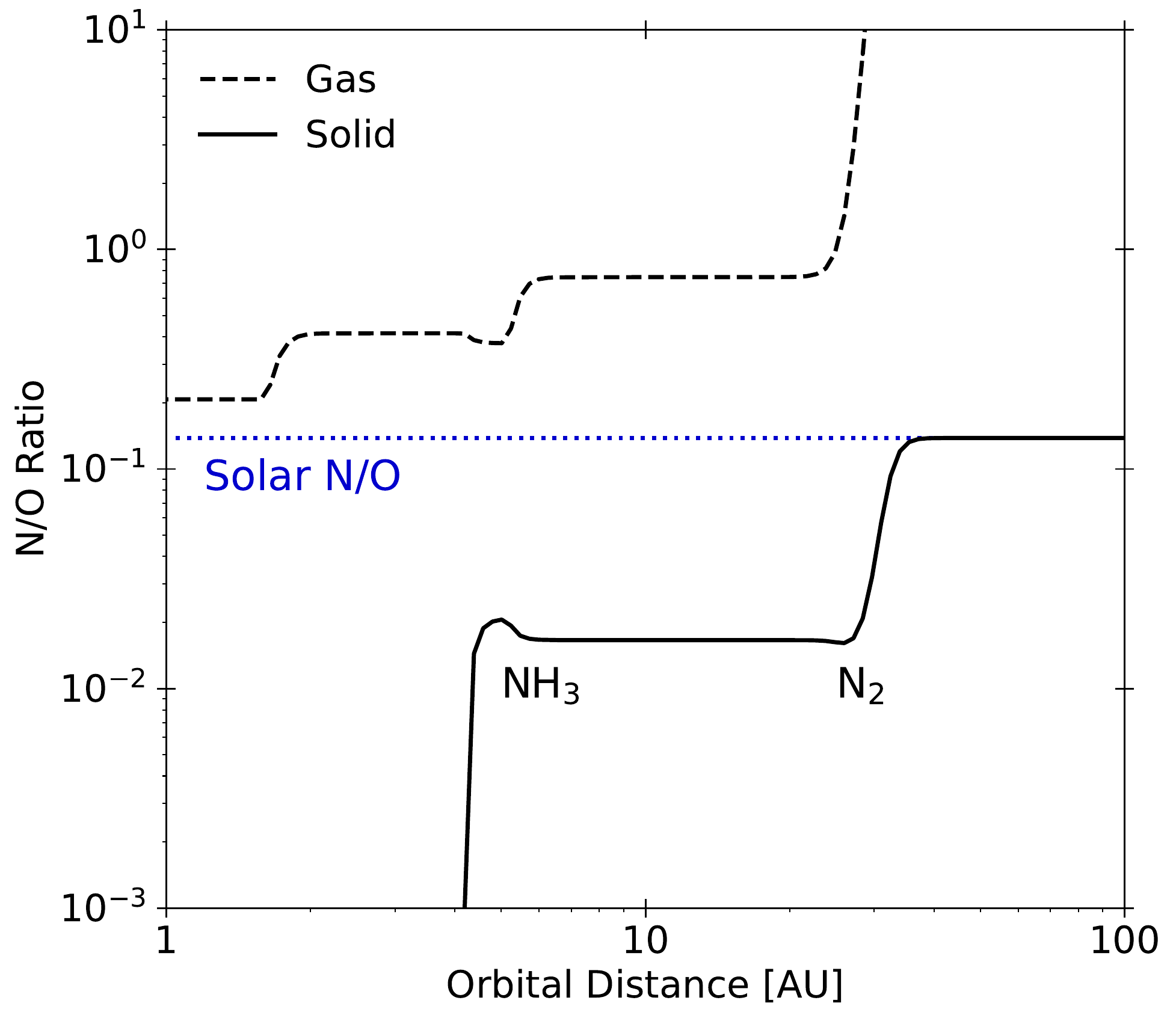}
\caption{The nitrogen-to-oxygen ratio as a function of orbital distance in a protoplanetary disk model. The black dashed and solid lines show the N/O ratio of disk gas, and solids, respectively.  The solar ratio is shown in dotted blue.  We compute this profile using the phase equilibrium model of \citet{Ohno&Ueda21} assuming the disk model of \citet{Oberg&Wordsworth19}.
}
\label{fig:NtoO}
\end{figure}

\rev{
To clarify the usefulness of the N/O ratio for example, Figure \ref{fig:NtoO} shows the nitrogen-to-oxygen ratio (N/O) of solids and gas in a protoplanetary disk computed by the phase equilibrium model of \citet{Ohno&Ueda21} for the protosolar disk model of \citet{Oberg&Wordsworth19}. %The figure demonstrates the monotonic increase in gas-phase N/O and an order-of-magnitude variation in solid-phase N/O.
The gas-phase N/O monotonically increases with orbital distance, as the O-bearing molecules (e.g., H$_2$O, CO$_2$) are gradually removed from gas phase via condensation while most of N remains as a highly volatile N$_2$ gas within N$_2$ snowline.
The solid-phase N/O shows an order-of-magnitude orbital variation because of large abundance difference between NH$_3$ and N$_2$ \citep[e.g.,][]{Oberg&Bergin21}.
The latter indicates the strong dependence of atmospheric N/O to formation location if solid (e.g., planetesimal) accretion predominantly determines the atmospheric composition.
}
%As discussed in \citet{Ohno&Ueda21}, the atmospheric N/O is expected to be sensitive to formation location if disk solids, such as pebbles and planetesimals, determine the atmospheric composition, as solid N/O ratio has an order-of-magnitude variation as a function of a radial distance. 
%\footnote{This large radial variation of N/O is caused by the fact that N$_2$ and NH$_3$, suggested to be the main nitrogen reservoirs in the disks, have an order-of-magnitude difference between their abundances \citep[e.g.,][]{Oberg&Bergin21}. }

%\citet{Ohno&Ueda21} also appealed 

%Thus, it is expected that nitrogen-involved elemental ratios, such as N/O, have orders-of-magnitude radial variations in protoplanetary disks \citep[see e.g.,][]{Ohno&Ueda21}.
%Nitrogen-rich atmospheres could indicate the planet formation at extremely cold environments because N$_2$, the most abundant nitrogen reservoir, could be condensed into ices only at temperature of $\la 30~{\rm K}$.
It is also worth noting that several recent studies have discussed formation scenarios for Jupiter in our solar system.
Motivated by the nitrogen abundance being comparable to other heavy elements in the Jovian atmosphere \citep[for recent review, see][]{Guillot+22,Atreya+22}, \citet{Oberg&Wordsworth19} and \citet{Bosman+19} proposed that Jupiter might have initially formed outward of the N$_2$ snowline beyond $30~{\rm AU}$ where solid elemental ratios coincide with solar values \citep[see also][]{Owen+99}.
\citet{Ohno&Ueda21} suggested that the Jovian atmospheric composition could also be explained if Jupiter formed at a locally cold disk region caused by the shadow cast by a disk substructure, such as a dust pileup at H$_2$O snowline, which is not nearly at far away from the central star. 
%While these studies assumed that disk solids determine the Jovian atmospheric compositions through e.g., planetesimal dissolution to the envelope \citep[e.g.,][]{Pollack+86,Shibata&Helled22}, several studies have also discussed the origin of the Jovian atmospheric composition from the perspective of the evolution of disk gas compositions \citep[][]{Guillot&Hueso06,Monga&Desch15,Ali-dib17,Mousis+19,Schneider&Bitsch21b,Aguichine+22}.
%Second, observable Oxygen abundances in atmospheres could inevitably deviate from the true bulk abundance owing to the sequestration into condensed minerals, such as Mg$_2$SiO$_4$ \citep{Helling+19}. 
%This is unlikely the case of Nitrogen.

%---------------------
%---------------------

%Although the Nitrogen potentially provides valuable insights on planet formation processes, atmospheric chemistry makes it a non-trivial task to retrieve the bulk nitrogen abundances in the atmosphere.
%While nitrogen potentially provides valuable insights on planet formation processes, it is not easy, however, to constrain the atmospheric nitrogen abundance.
In substellar atmospheres, in the absence of ionizing flux, N$_2$ and NH$_3$, are the main nitrogen reservoirs \citep{Lodders&Fegley02}. 
HCN can also be abundant if photochemical processes are at work \citep{Moses+13}. 
NH$_3$ and HCN would be likely detectable by near future infrared observations by JWST and Ariel \citep{MacDonald&Madhusudhan17}, while N$_2$ is in general not observable due to \rev{the negligibly low visible and infrared opacity for the temperature regime of exoplanets.} %\footnote{N$_2$ actually has moderate opacity at extremely high temperature ($>4000~{\rm K}$). Perhaps, high-resolution spectroscopy might have a chance to detect N$_2$ if it abundantly exists in hot thermosphere.}}.
%the lack of visible and infrared spectral features for N$_2$.
\citet{Hobbs+19} used a photochemical kinetic model to show that the abundances of C- and O-bearing species, such as H$_2$O and CO, are insensitive to N/H ratio in hot Jupiters like HD 209458b.
%\citet{Hobbs+21} suggested that observable NH$_3$ and HCN abundances in hot Jupiter with equilibrium temperature of $T_{\rm eq}=1000$ and $2000~{\rm K}$ may be less sensitive to formation location than Oxygen- and Carbon-bearing species are.
\citet{Ramirez+20} also investigated the impact of N/H ratio on TiO abundances in ultra-hot Jupiters and found that the TiO abundance is nearly independent of N/H.  
Since C- and O-bearing species abundances are insensitive to bulk nitrogen abundance \rev{in sub-stellar atmospheres}, it appears that the only route to diagnose \rev{the bulk nitrogen abundance of a giant planet is from NH$_3$ and/or HCN.}%a planet's bulk nitrogen abundance is from NH$_3$ and/or HCN.
%demonstrating the difficulty of constraining N abundance from TiO detection.
%Since N$_2$ is effectively invisible due to the lack of visible and infrared spectral features, only NH$_3$ and HCN are accessible nitrogen species.
%\citet{MacDonald&Madhusudhan17} showed that noticeable spectral features exist at $\sim2.2~{\rm {\mu}m}$ for NH$_3$ and $\sim3.1$ and $4.0~{\rm {\mu}m}$ for HCN, which makes the detection of nitrogen-bearing species viable from high precision observations at $1$--$5~\rm{\mu}m$ by JWST and Ariel.

%On the other hand, it is not straightforward to constrain the bulk nitrogen abundance from NH$_3$ and HCN.
%As we will demonstrate
\rev{However}, constraining the bulk nitrogen abundance from NH$_3$ and HCN is a complex task.
The NH$_3$ and HCN abundances in the observable atmosphere readily deviates from thermochemical equilibrium abundances because of disequilibrium effects, such as vertical mixing and photochemistry \citep[e.g.,][]{Moses+11,Line+11,Venot+13}. 
%demonstrated that NH$_3$ and HCN abundances readily deviate from the thermochemical equilibrium in HD189733b and HD209458b atmospheres owing to vertical mixing and photochemistry (see also \citealt{Venot+13}).
For warm planets of $T_{\rm eq}\la1000~{\rm K}$, \citet{Fortney+20} investigated disequilibrium NH$_3$ abundances on Saturn-like planets with various $T_{\rm eq}$ and found that NH$_3$ abundance depends on a number of factors, such as planetary mass, age, and metallicity.
They also suggested that N$_2$ will actually dominate over NH$_3$ over a very wide range of temperature and ages, making the observable NH$_3$ abundance only a lower limit of bulk nitrogen abundance.
\citet{Hu21} also investigated photochemistry on temperate/cold H$_2$-rich planets and found that NH$_3$ tends to be depleted due to photodissociation, especially on planets around G/K stars.

In this study, we expand the work of \citet{Fortney+20} with a particular focus on nitrogen chemistry.
\rev{Here in Paper I we systematically investigate the thermal structure of planetary deep atmosphere, which significantly affects the disequilibrium abundance of NH$_3$, as demonstrated by \citet{Fortney+20}.}
%We systematically investigate the relationship between observable NH$_3$ and HCN abundances and bulk nitrogen abundance using both semi-analytical arguments and photochemical kinetic models.
%Then, we will identify the phase space of planetary properties with which NH$_3$ abundance would reflect the bulk nitrogen abundance.
\rev{While \citet{Fortney+20} investigated the effects of planetary deep atmospheres using numerical models, this study advances the field by establishing a semi-analytical model that explicitly links planetary gravity, intrinsic temperature, metallicity, bulk nitrogen abundance, and disequilibrium NH$_3$ abundance. The model is readily applicable to arbitrarily planets and will be useful to interpret the retrieved NH$_3$ abundance in future observations.} 
%we attempt to provide a simple diagnostic model that provides a first-order constrain on the bulk nitrogen abundance from the retrieved NH$_3$ abundance.
%We will identify a planetary mass-age space in which the NH$_3$ abundance diagnoses the bulk nitrogen abundance based on a semi-analytical argument.

The organization of this paper is as follows.
%Then, we verify our classification scheme using a detailed disequilibrium photochemical model, which is significantly improved from \citet{Fortney+20} who relied on the vertical quench approximation based solely on timescale argument.
In Section \ref{sec:overview}, we introduce a basic background of nitrogen equilibrium and disequilibrium chemistry.
In Section \ref{sec:deep_adiabat}, we investigate atmospheric pressure-temperature (\emph{P--T}) profiles for a wide range of planetary parameters.  We derive a semi-analytical fit to understand why giant planets typically have a universal deep adiabat, irrespective of incident flux, which has a major impact on NH$_3$ abundances from disequilibrium chemistry from vertical mixing. %\footnote{We refer the equilibrium temperature to the temperature for zero Bond albedo with full heat redistribution unless otherwise indicated.}.
In Section \ref{sec:N_map}, we identify the relation between NH$_3$ and bulk nitrogen abundances as a function of planetary parameters from semi-analytical arguments.
%Then, we verify our predictions by applying a photochemical kinetics model to a wide range of planetary parameters.
In Section \ref{sec:discussion}, we \rev{describe caveats of this study.} %discuss the effects of stellar spectral type and the UV-shielding by photochemical hazes.
%In particular, we examine the UV-shielding effects of haze on atmospheric photochemistry by implementing a physically motivated haze model to the photochemical kinetic model.
In Section \ref{sec:summary}, we summarize our findings.
\rev{In the paper II of this series \citep{Ohno&Fortney22b}, we verify our semi-analytical predictions using a photochemical kinetics model and discuss the observational implications for atmospheric nitrogen species on transmission and emission spectra.}% to a wide range of planetary parameters.

%%%%%%%%%%%%%%%%%%%%%%%%%%%%%%%%%
\section{Nitrogen Chemistry: The Importance of The Deep Atmosphere Structure}\label{sec:overview}

One of the important factors in controlling the observable NH$_3$ abundance is vertical  vertical mixing within an atmosphere.
Atmospheric compositions follow thermochemical equilibrium in the deep hot atmospheres, while the abundances at lower pressure, where it is colder, tend to be out of equilibrium, vertically constant, and reflect the equilibrium compositions of the deep atmosphere \rev{(though see Section \ref{sec:caveat} for a caveat on this picture)}. 
This phenomena if often called ``quenching'' \citep[e.g.,][]{Fegley&Prinn85,Fegley&Lodders94,Zahnle+14,Tsai+18} and was originally identified for CO/CH$_4$ in the Jovian atmosphere, where detected CO abundances are many orders of magnitude higher than thermochemical equilibrium calculations \citep[][]{Prinn&Barshay77}.
This quenching is caused by the slow thermochemical conversion, compared to relatively fast vertical mixing \citep[e.g.,][]{Moses+11}.
Several recent studies have attempted to constrain the strength of vertical mixing in brown dwarf and giant planet atmospheres from quenched molecular abundances \citep{Miles+20,Kawashima&Min21,Mukherjee+22b}.

%, as air parcel moves before thermochemical conversion takes place \citep[see e.g.,][]{Moses+11}.
%\subsection{Overview}
%Our approach for constraining bulk nitrogen abundance is straightforward: we will attempt to recover the bulk nitrogen abundance from NH$_3$ whose abundance is accessible by spectroscopic observations.
%The most ideal situation is that NH$_3$ accommodates all nitrogen, yielding NH$_3$ abundance identical to the bulk nitrogen abundance.
%In reality, a fraction of nitrogen would be sequestered to N$_2$ that cannot be detected by spectroscopic observations.
%Taking these facts into consideration, we attempt to provide a prescription with which one can convert the observationally constrained NH$_3$ abundance to bulk nitrogen abundance for given planetary properties, such as mass and age.

Since the upper atmospheric composition is related to the composition of deep hot atmosphere, it is necessary to understand the planetary deep atmosphere and interior to relate the observable NH$_3$ abundance with bulk nitrogen abundance \citep{Fortney+20}.
To this end, we first introduce the nitrogen chemistry in deep atmospheres where thermochemical equilibrium is expected.
%Subsequently, we discuss the possible presence of a common deep adiabatic profile for irradiated substellar atmospheres from analytical arguments in Section \ref{sec:deep_adiabat}.
%Then, we investigate how the various planetary parameters controls the thermal structure of deep atmosphere using a radiative-convective equilibrium model in Section \ref{sec:RCE_result}.

\subsection{Thermochemical equilibrium and vertical quenching of NH$_3$}\label{sec:nitrogen_chem}
%%%%%%%%%%%%%%%%%%%%%%%%%%%%%
\begin{figure*}[t]
\centering
\includegraphics[clip, width=\hsize]{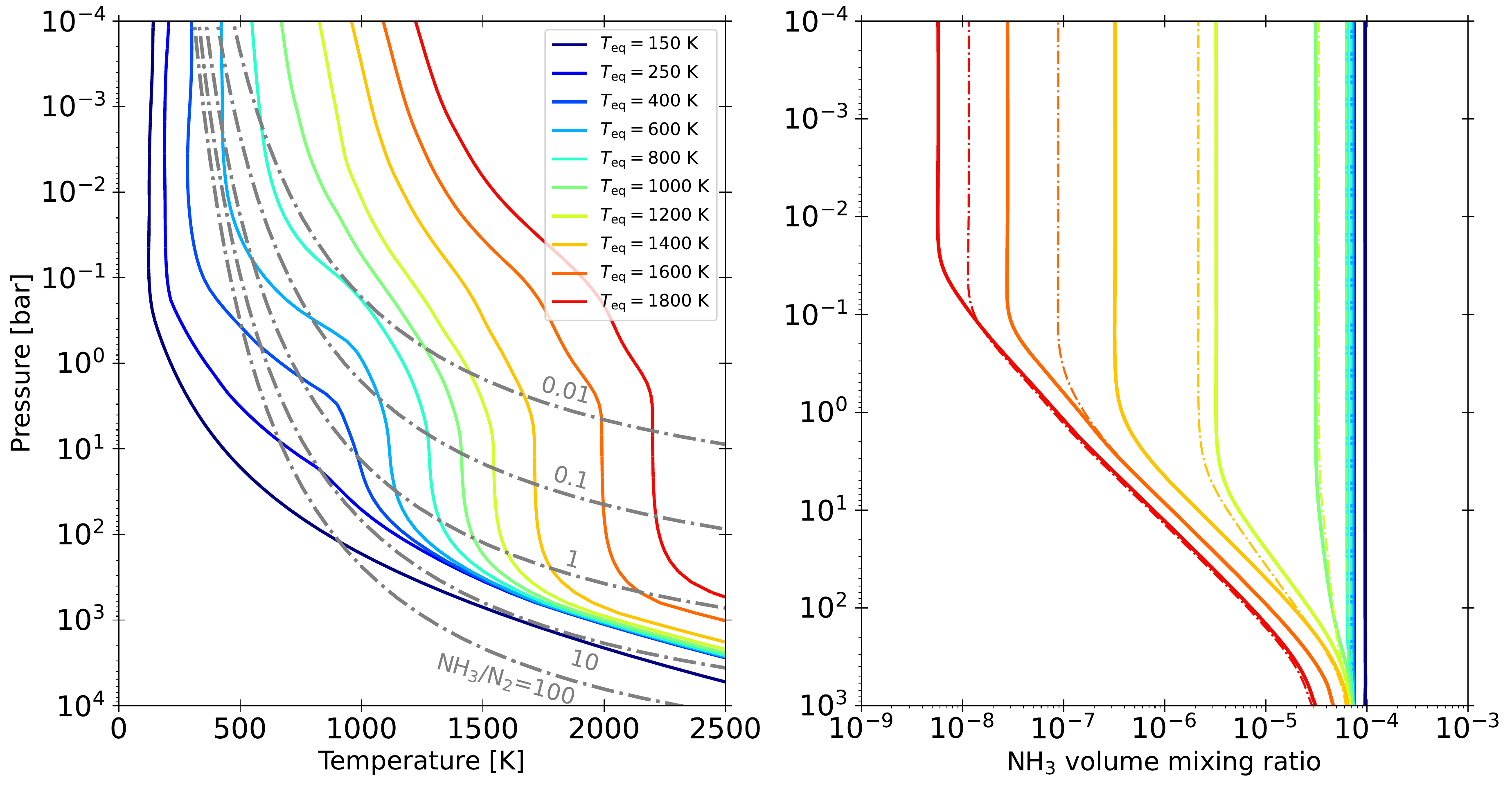}
\caption{(Left) \emph{P--T} profiles of solar composition atmospheres computed in radiative-convective equilibrium (see Section \ref{sec:RCE_result} for details). Different colored lines show the \emph{P--T} profiles for different equilibrium temperatures. We assume a surface gravity of $g=10~{\rm m~s^{-2}}$ and planetary intrinsic temperature of $T_{\rm int}=100~{\rm K}$. The gray dashed lines show the abundance ratio contours of ${\rm NH_3/N_2}=0.01$, $0.1$, $1$, $10$, $100$ from top to bottom, computed by Equation \eqref{eq:contour_N2_NH3_1}. \rev{(Right) Vertical distribution of NH$_3$ volume mixing ratio for the \emph{P--T} profiles in the left panel. The solid and dash-dot lines show the distributions for eddy diffusion coefficients of $10^{8}$ and ${10}^{10}~{\rm cm^2~s^{-1}}$, respectively. Note that the distributions of different eddy diffusion coefficients are almost superimposed on each other at $T_{\rm eq}<1000~{\rm K}$. We have turned off photochemistry for the sake of simplicity.}}
\label{fig:PT_example}
\end{figure*}
%%%%%%%%%%%%%%%%%%%%%%%%%%%%%
%We first consider NH$_3$ abundance controlled by vertical mixing, the so-called vertical quenching \citep[e.g.,][]{Prinn&Barshay77,Fegley&Prinn85,Fegley&Lodders94}, and neglect the effect of photochemistry for this moment.
%The upper atmosphere composition is then determined by the equilibrium abundance at a certain level of deep atmospheres where the chemical interconversion timescale becomes slower than the vertical mixing timescale.
%Therefore, 

The quenching behavior of NH$_3$ has an interesting characteristic: the quenched NH$_3$ abundance is insensitive to the strength of vertical mixing \citep{Saumon+06,Zahnle+14,Fortney+20}.
As discussed in \citet{Zahnle+14}, this is caused by the abundance ratio contours of $\rm NH_3/N_2$ being nearly parallel to the adiabatic profiles of substellar atmospheres (see also Figure \ref{fig:PT_example}). 
The vertically quenched abundance is determined, to a good approximation, by the equilibrium abundance at certain depth where thermochemical interconversion timescale becomes equal to the vertical mixing timescale.
However, since the deep adiabat is nearly along the contour of constant $\rm NH_3/N_2$ ratio, the quenched NH$_3$ abundance is nearly the same wherever the quenching takes place.
This characteristic has an advantage in interpreting the quenched NH$_3$: one does not need to worry too much about the uncertainty of vertical mixing strength, parameterized by $K_{\rm zz}$.

\rev{To further clarify the quenching behavior of NH$_3$, the right panel of Figure \ref{fig:PT_example} shows the vertical distribution of NH$_3$ in a solar composition atmosphere computed by the chemical kinetics code VULCAN \citep{Tsai+17,Tsai+21} for various planetary equilibrium temperature and eddy diffusion coefficients. While the NH$_3$ distribution in the upper atmosphere depends on the eddy diffusion coefficient for hot  ($T_{\rm eq}\ga1000~{\rm K}$) planets where the quenching occurs at shallower radiative parts of the atmosphere, the abundances are nearly independent of the eddy diffusion at warm ($T_{\rm eq}<1000~{\rm K}$) planets where the quenching occurs at deep adiabatic atmospheres.}

%If there are constrain on the property of the deep adiabat, we could infer the bulk nitrogen abundance from the quenched NH$_3$ abundance.
Since the quenched abundance is determined by the composition of a deep atmosphere point where thermochemical equilibrium is valid, it is worth first examining the equilibrium abundance of nitrogen species.
The law of mass action provides the relation of N$_2$ and NH$_3$ that should be satisfied in thermochemical equilibrium, given by
\begin{equation}\label{eq:nitrogen_eq}
    \frac{P_{NH_3}^2}{P_{N_2}P_{H_2}^3}=K_{\rm N_2 \rightleftharpoons NH_3}=Ae^{B/T},
\end{equation}
where $P_{\rm H_2}$, $P_{\rm N_2}$, and $P_{\rm NH_3}$ are the partial pressure of H$_2$, N$_2$, and NH$_3$, respectively, $K_{\rm N_2 \rightleftharpoons NH_3}$ is the equilibrium constant of N$_2$-NH$_3$ interconversion (N$_2$+3H$_{\rm 2}$ $\rightleftharpoons$2NH$_3$), $A=5.90\times{10}^{-13}~{\rm bar^{-2}}$, and $B=13207~{\rm K}$ \citep{Zahnle+14}.
Assuming that N$_2$ and NH$_3$ accommodate most of the nitrogen, we can approximate the nitrogen conservation as $f_{\rm N}\approx f_{\rm NH_3}+2f_{\rm N_2}$, where $f_{\rm N}$, $f_{\rm N_2}$, and $f_{\rm NH_3}$ are volume mixing ratios of total nitrogen, N$_2$, and NH$_3$.
Then, one can obtain the pressure-temperature relation of $\rm NH_3/N_2=\xi$ contour as \citep{Zahnle+14}
\begin{equation}\label{eq:contour_N2_NH3_1}
     f_{\rm N}\left( \frac{\xi^2}{2+\xi }\right) =P^2 f_{\rm H_2}^3Ae^{B/T},
\end{equation}
%This equation indicates that NH$_3$ abundance is proportional to $f_{\rm N}^{1/2}$ for N$_2$ dominated regime (i.e., $\xi \ll 1$), which may explain why NH$_3$ abundance is relatively insensitive to planet formation location in the hot Jupiter simulations of \citet{Hobbs+21}.
where $f_{\rm H2}={\rm H_2/(H_2+He)}=0.859$ and $f_{\rm N}=1.16\times{10}^{-4}$ in solar elemental abundances of \citet{Asplund+21}.
We note that $f_{\rm N}$ is not identical to the N/H ratio, as it is given by
\begin{equation}\label{eq:N/H}
    f_{\rm N}={\rm \frac{N}{H_2+He } }={\rm \frac{2N/H}{ 1+2He/H } }=2f_{\rm H_2}{\rm N/H},
\end{equation}
where $\rm N/H=6.76\times{10}^{-5}$ is the value for solar composition \citep{Asplund+21}.
Equation \eqref{eq:contour_N2_NH3_1} is inconvenient from an observational perspective, as both $f_{\rm N}$ and $\xi$ are unknown.
%One can instead eliminate the $f_{\rm N_2}$ in Equation \eqref{eq:nitrogen_eq} using $f_{\rm N}\approx f_{\rm NH_3}+2f_{\rm N_2}$ relation, yielding
Instead, eliminating $f_{\rm N_2}$ in Equation \eqref{eq:nitrogen_eq} using $f_{\rm N}\approx f_{\rm NH_3}+2f_{\rm N_2}$, we obtain
\begin{equation}\label{eq:f_N1}
    {\rm N/H}=\frac{f_{\rm NH_3}}{2f_{\rm H_2}}\left[ 1 + \frac{2f_{\rm NH_3}e^{-B/T}}{Af_{\rm H_2}^3P^2} \right],
\end{equation}
where we use Equation \eqref{eq:N/H}.
Under chemical equilibrium, Equation \eqref{eq:f_N1} can straightforwardly constrain the bulk nitrogen abundance from the NH$_3$ abundance.
In addition, as introduced above, the equilibrium NH$_3$ abundance is approximately constant along the deep adiabatic profiles.
Thus, it is expected that the quenched NH$_3$ abundance is mostly determined by the deep adiabatic profile alone.

%%%%%%%%%%%%%%%%%%%%%%%%%%%%%
\section{Constraining Thermal Structures of Deep Atmospheres}\label{sec:deep_adiabat}
\subsection{Radiative Zero Solution of Irradiated Exoplanets}
%As discussed in previous section, constraining the PT relation of the deep adiabat is crucial to retrieve bulk nitrogen abundance from observable NH$_3$ abundance.
The preceding argument highlights the importance of identifying the thermal structures of deep atmospheres (below the photosphere) to relate the quenched NH$_3$ abundance with bulk nitrogen abundance. 
Here, we point out an interesting trend of deep atmospheres: many planets with different equilibrium temperatures\footnote{We refer the equilibrium temperature to the temperature for zero Bond albedo with full heat redistribution unless otherwise indicated.} ($T_{\rm eq}=250$--$1200~{\rm K}$) have nearly the same deep adiabatic profile as seen in Figure \ref{fig:PT_example}.
\citet{Fortney+07} first reported such a universal deep adiabat in their radiative-convective models.
Motivated by simplified calculations with dual-band radiative transfer \citep{Guillot10}, \citet{Fortney+20} speculated that the universal deep adiabat may emerge owing to the steep change of visible-to-infrared opacity ratio caused by the loss of gas-phase alkali metals, although the actual cause still remains unclear.
\rev{The universality of the deep adiabat has a crucial impact on the disequilibrium abundance of NH$_3$: the quenched NH$_3$ abundance is nearly independent of the equilibrium temperature for temperate to warm exoplanets, as seen in the right panel of Figure \ref{fig:PT_example}.}

Here, we elaborate why many planets have nearly the same deep adiabatic profile for a wide range of equilibrium temperatures.
The common thermal structure independent of upper boundary conditions is reminiscent of the ``radiative zero solution'' discussed in the context of stellar and protoplanetary envelope structures \citep[e.g.,][]{Hayashi+62,Mizuno80,Stevenson+82,Kippenhahn+94}.
In purely radiative atmospheres without convection, the atmospheric temperature structure follows
\begin{eqnarray}\label{eq:nabla_rad2}
%\nonumber
    \left( \frac{d\ln T}{d\ln P}\right)_{\rm rad}&=&\frac{3\kappa L_{\rm int}}{64\pi \sigma GM}\frac{P}{T^4}=\frac{3\kappa_{\rm 0} P^{1+\alpha}T^{\rm \beta}}{16 g}\frac{T_{\rm int}^4}{T^4},
\end{eqnarray}
where $L_{\rm int}=4\pi R^2 \sigma T_{\rm int}^4$ is the planetary intrinsic luminosity, $T_{\rm int}$ is the planetary intrinsic temperature, $\sigma$ is the Stefan-Boltzmann constant, $g=GM/R^2$ is the planetary gravity, and $\kappa=\kappa_{\rm 0}P^{\alpha}T^{\beta}$ is the atmospheric Rosseland-mean opacity.
Assuming constant gravity, $\alpha$, and $\beta$, Equation \eqref{eq:nabla_rad2} yields an analytical solution of
\begin{equation}\label{eq:rad_zero}
    T=T_{\rm 0}+\left[ \frac{3\kappa_{\rm 0}T_{\rm int}^4(4-\beta)}{16g(1+\alpha)}\right]^{1/(4-\beta)} (P^{(1+\alpha)/(4-\beta)}-P_{\rm 0}^{(1+\alpha)/(4-\beta)}),
\end{equation}
where $P_{\rm 0}$ and $T_{\rm 0}$ are the pressure and temperature of the upper boundary.
For $(4-\beta)/(1+\alpha)>0$ \footnote{For $(4-\beta)/(1+\alpha)<0$, the temperature structure converges to the following isothermal profile in the limit of $P\gg P_{\rm 0}$: 
\begin{equation}
    \nonumber
    T\approx T_{\rm 0}-\left[ \frac{3\kappa_{\rm 0}T_{\rm int}^4P_{\rm 0}^{1+\alpha}(4-\beta)}{16g(1+\alpha)}\right]^{1/(4-\beta)}.
\end{equation}
In this case, the upper boundary conditions controls the temperature structure of the deep atmosphere.
}, since $P\gg P_{\rm 0}$ and $T\gg T_{\rm 0}$ in the limit of a deep atmosphere, the temperature structure asymptotically approaches the same temperature structure relation with $P_{\rm 0}=0$ and $T_{\rm 0}=0$ in Equation \eqref{eq:rad_zero} regardless of the upper boundary condition, which is called the radiative-zero solution \citep[e.g.,][]{Hayashi+62,Mizuno80,Stevenson+82,Kippenhahn+94}.
%Solving the equality of the first and second terms in Equation \eqref{eq:rad_zero} with respect to $P$ for $P_{\rm 0}=0$, we can evaluate a threshold pressure level below which the thermal structure forgets the upper atmospheric temperature as
%\begin{equation}
%    P_{\rm rad*}=\left[ \frac{16gT_{\rm 0}^{4-\beta}(1+\alpha)}{3\kappa_{\rm 0}T_{\rm int}^4(4-\beta)} \right]^{1/(1+\alpha)}.
%\end{equation}
%%%%%%%%%%%%%%%%%%%%%%%%%%%%%
\begin{figure*}[t]
\centering
\includegraphics[clip, width=0.49\hsize]{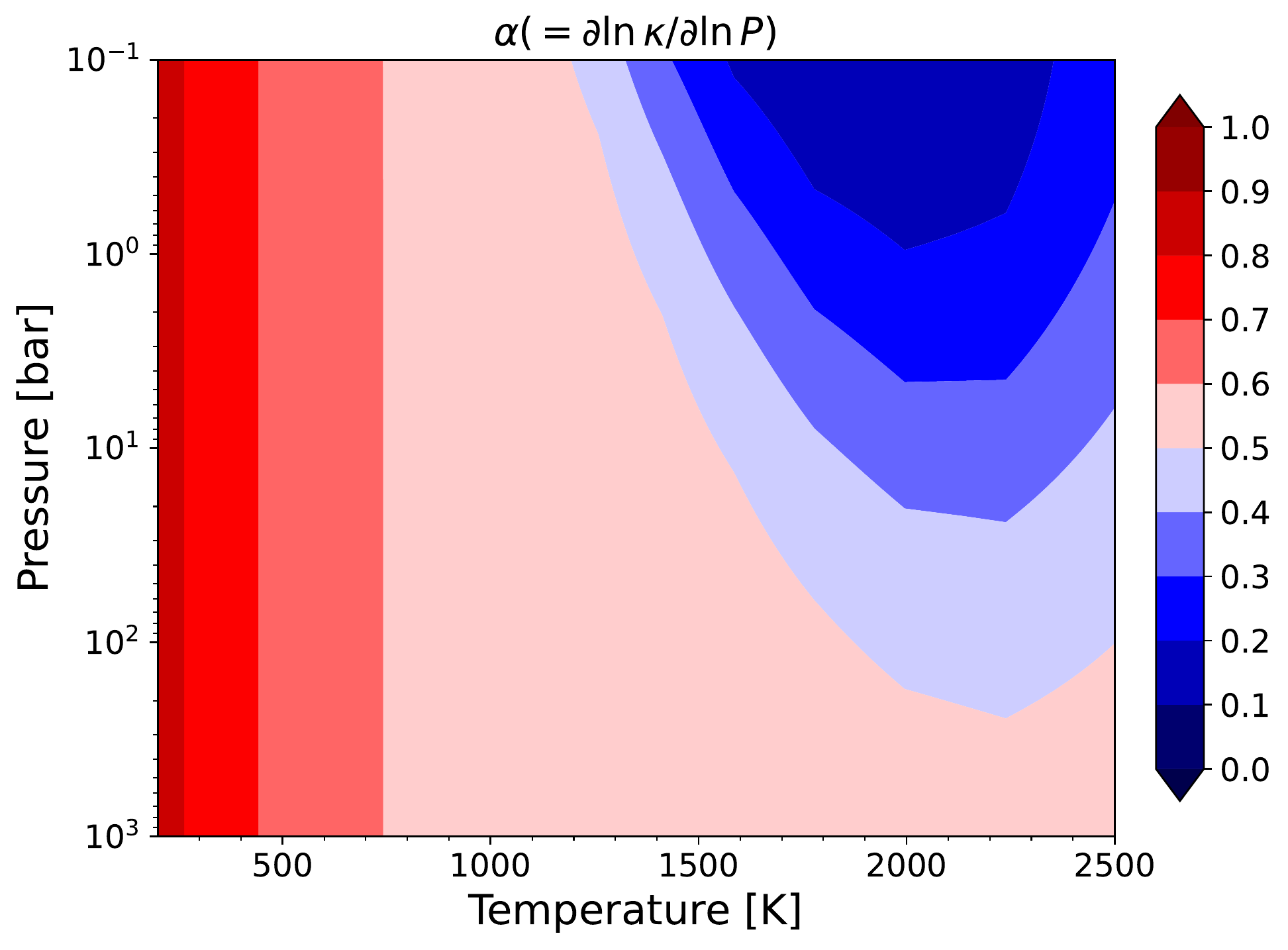}
\includegraphics[clip, width=0.49\hsize]{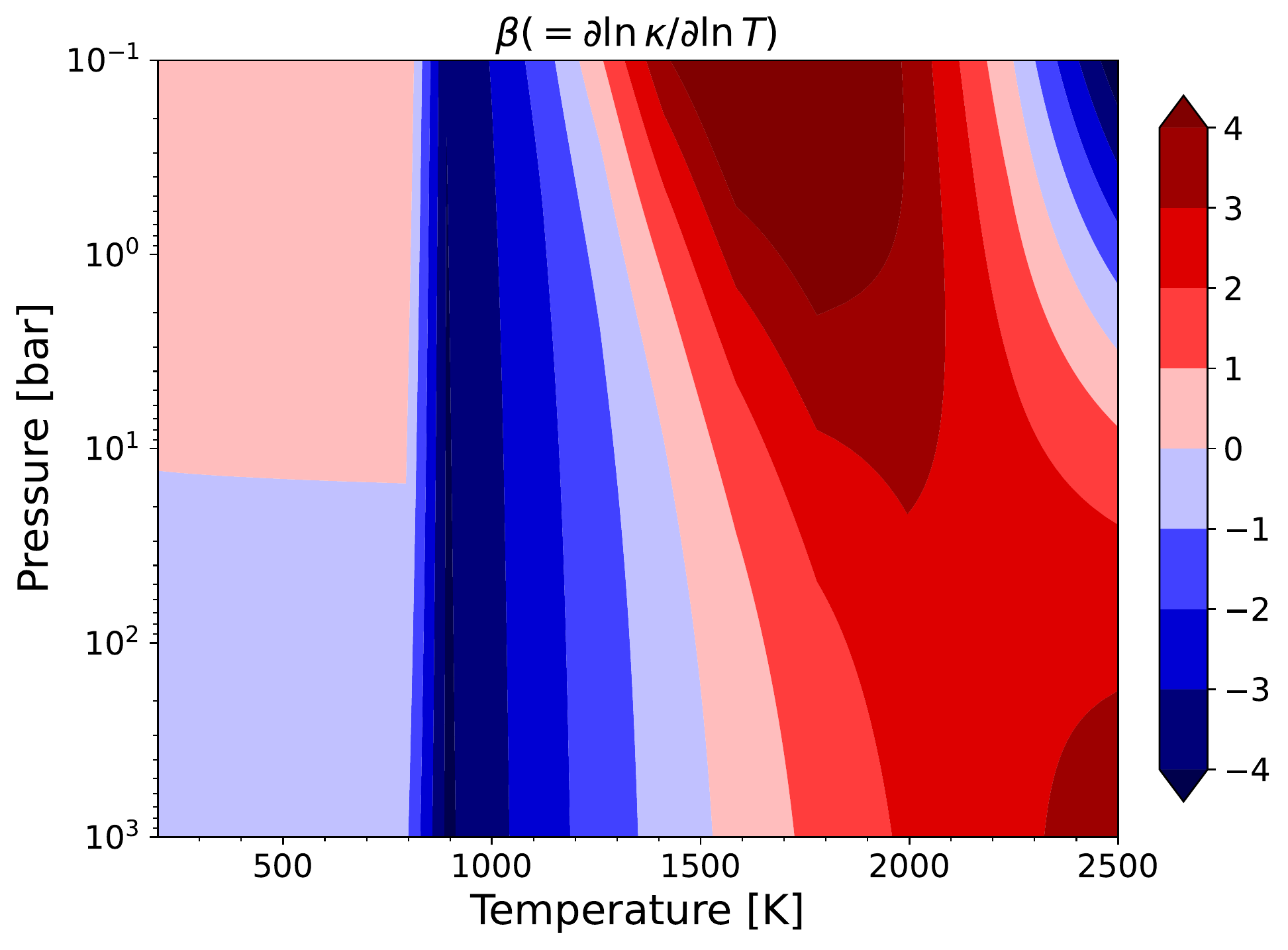}
\includegraphics[clip, width=0.49\hsize]{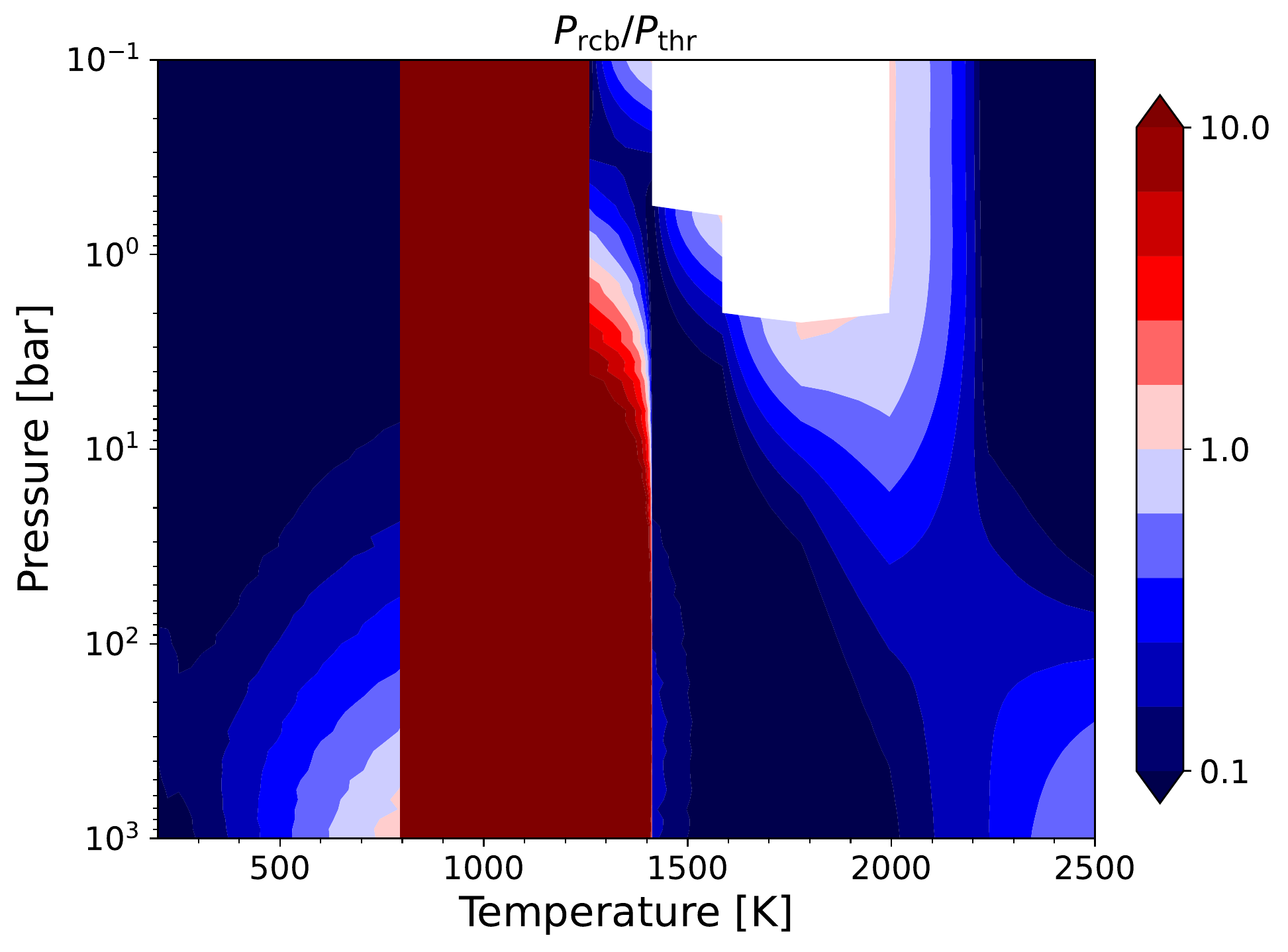}
\includegraphics[clip, width=0.49\hsize]{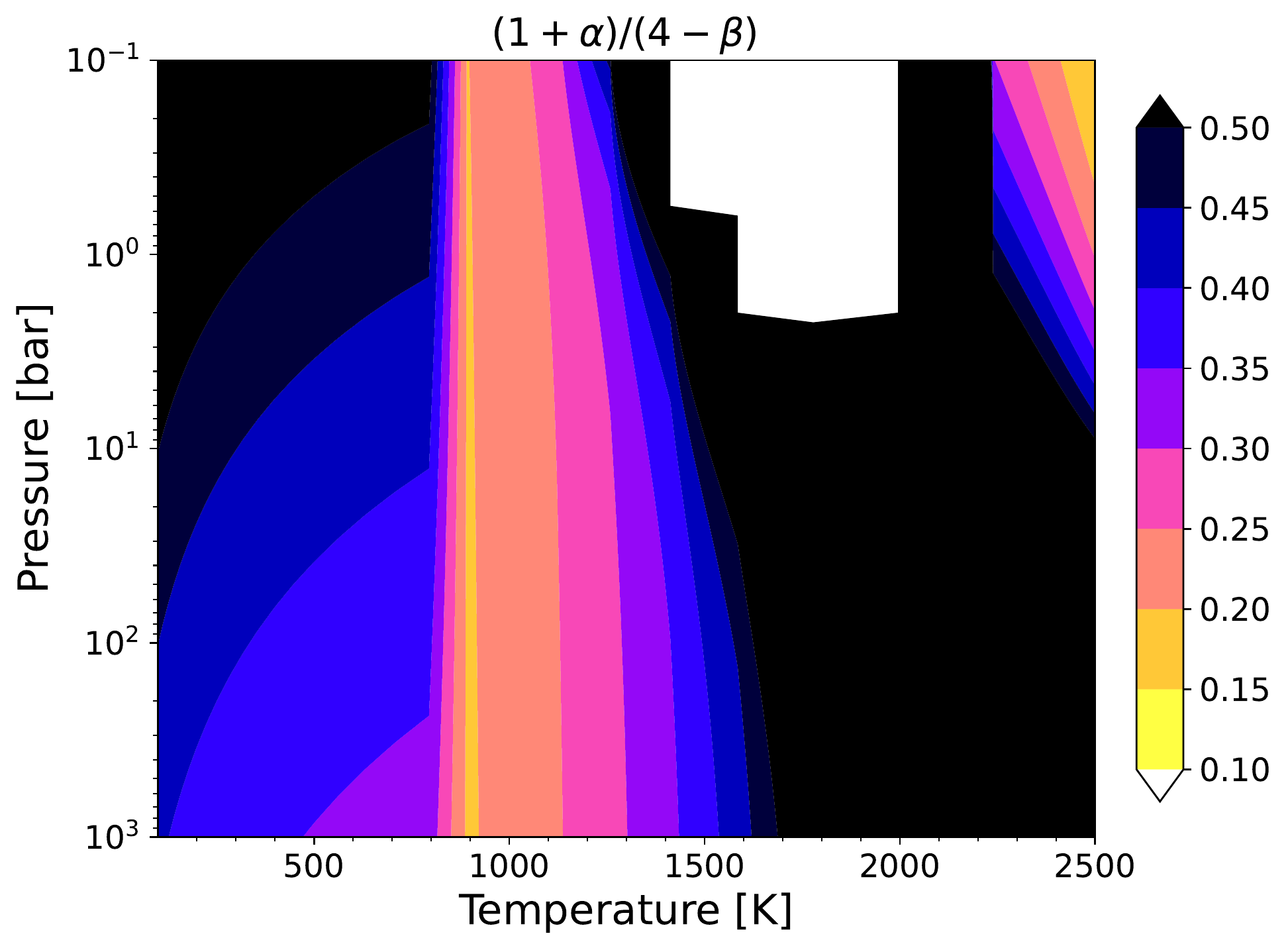}
\caption{(Upper left and right panels) Pressure and temperature dependence of the Rosseland mean opacity of a solar composition gas, based on the \citet{Freedman+14} analytic opacity fit.
(Lower left) The ratio of the RCB pressure to the threshold RCB pressure estimated by Equation \eqref{eq:metric_adiabat}. The \emph{P--T} profile converges to the radiative zero solution before it meets the RCB for $P_{\rm rcb}/P_{\rm thr}>1$ (redder colors), leading to the deep adiabatic profile that is insensitive to stellar insolation. The black contour denotes $P_{\rm rcb}/P_{\rm thr}=1$.
(Lower right) The pressure dependence of the radiative zero solution, $(1+\alpha)/(4-\beta)$. It is expected that the \emph{P--T} profiles tend to converge to the same radiative zero solution of $T\propto P^{(1+\alpha)/(4-\beta)}$ in the \emph{P--T} space with the same value of $(1+\alpha)/(4-\beta)$. We have filled the space of $(1+\alpha)/(4-\beta)<0$ in white for clarity. \rev{In computing the bottom left panel, we set $\phi=1$ when $\phi$ exceeds unity, as the \emph{P--T} profile always converges to the radiative-zero solution at $\phi\ge1$.}
%because the absence of convection effectively means $P_{\rm rcb}= \infty$.}.
}
\label{fig:dependence}
\end{figure*}
%%%%%%%%%%%%%%%%%%%%%%%%%%%%%
%%%%%%%%%%%%%%%%%%%%%%%%%%%%%
\begin{figure*}[t]
\centering
\includegraphics[clip, width=0.4\hsize]{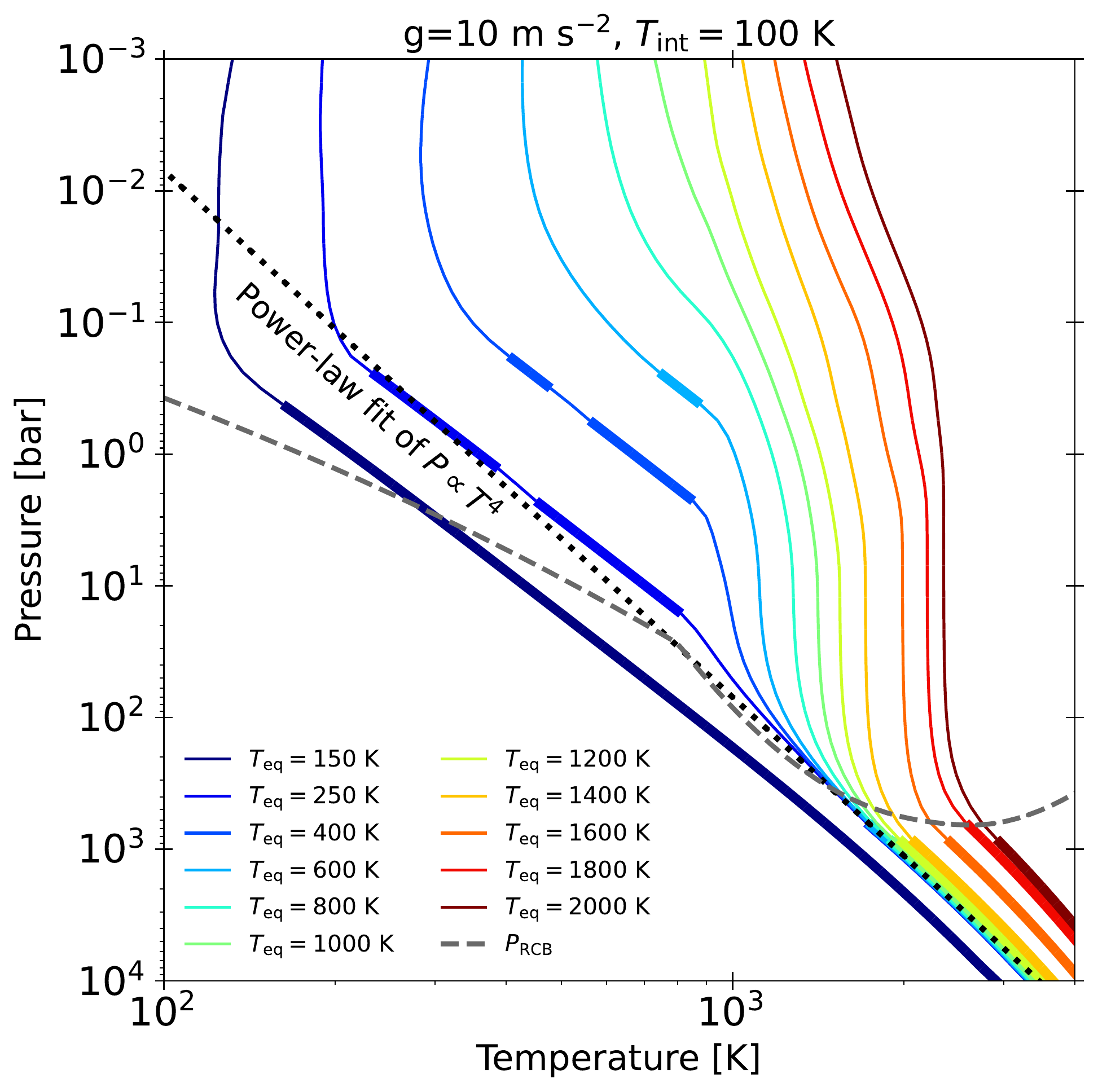}
\includegraphics[clip, width=0.4\hsize]{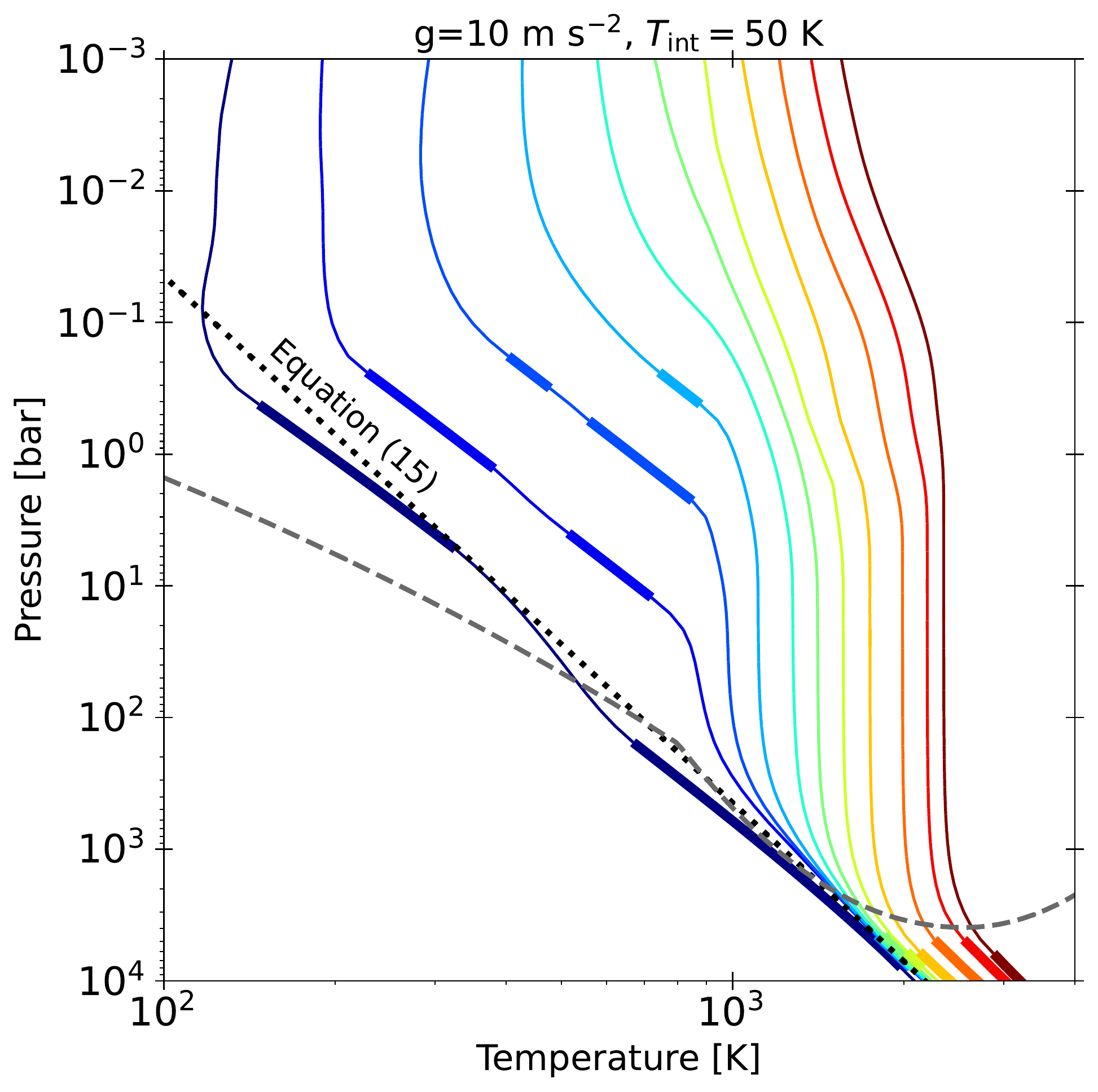}
\includegraphics[clip, width=0.33\hsize]{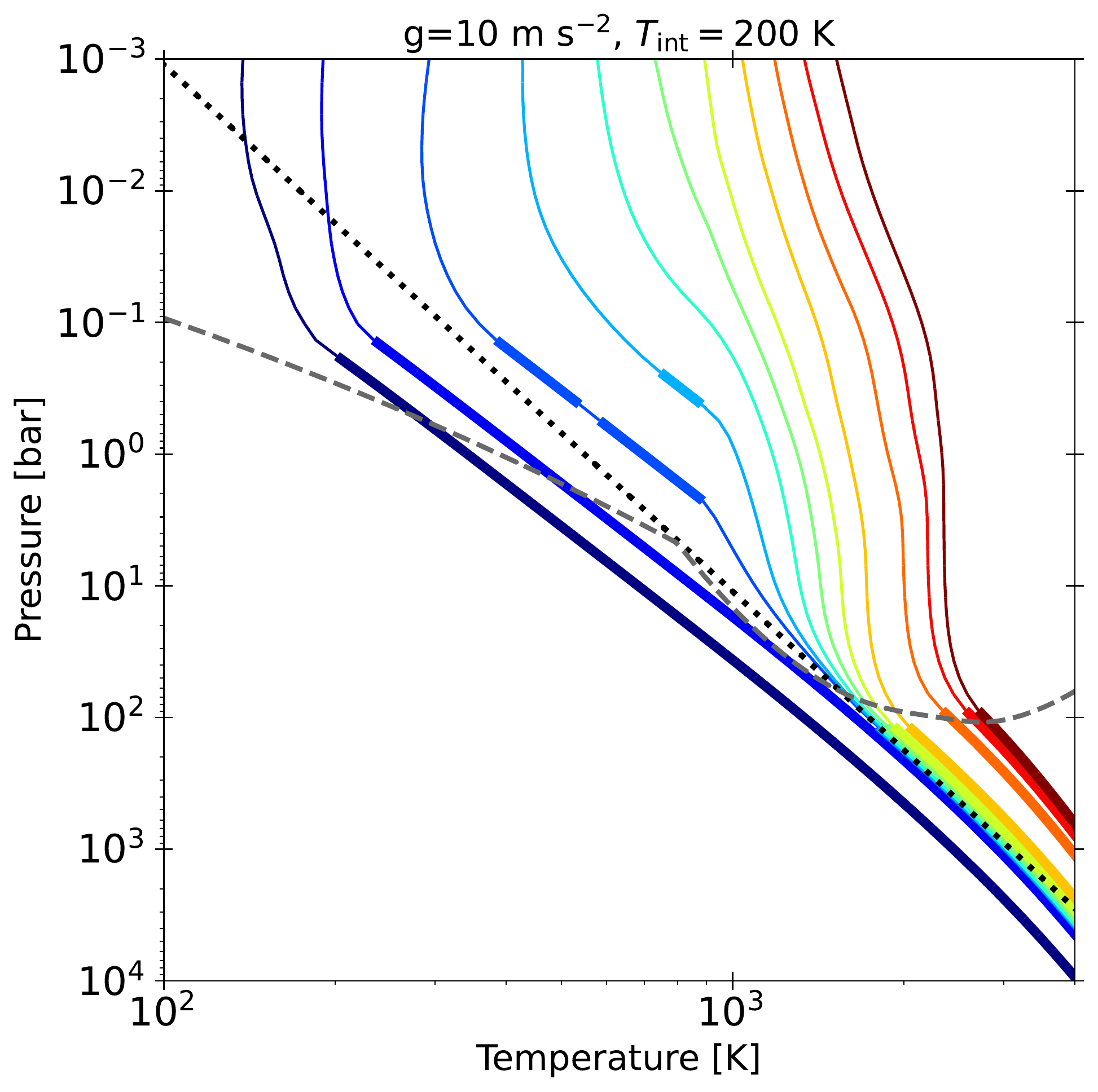}
\includegraphics[clip, width=0.33\hsize]{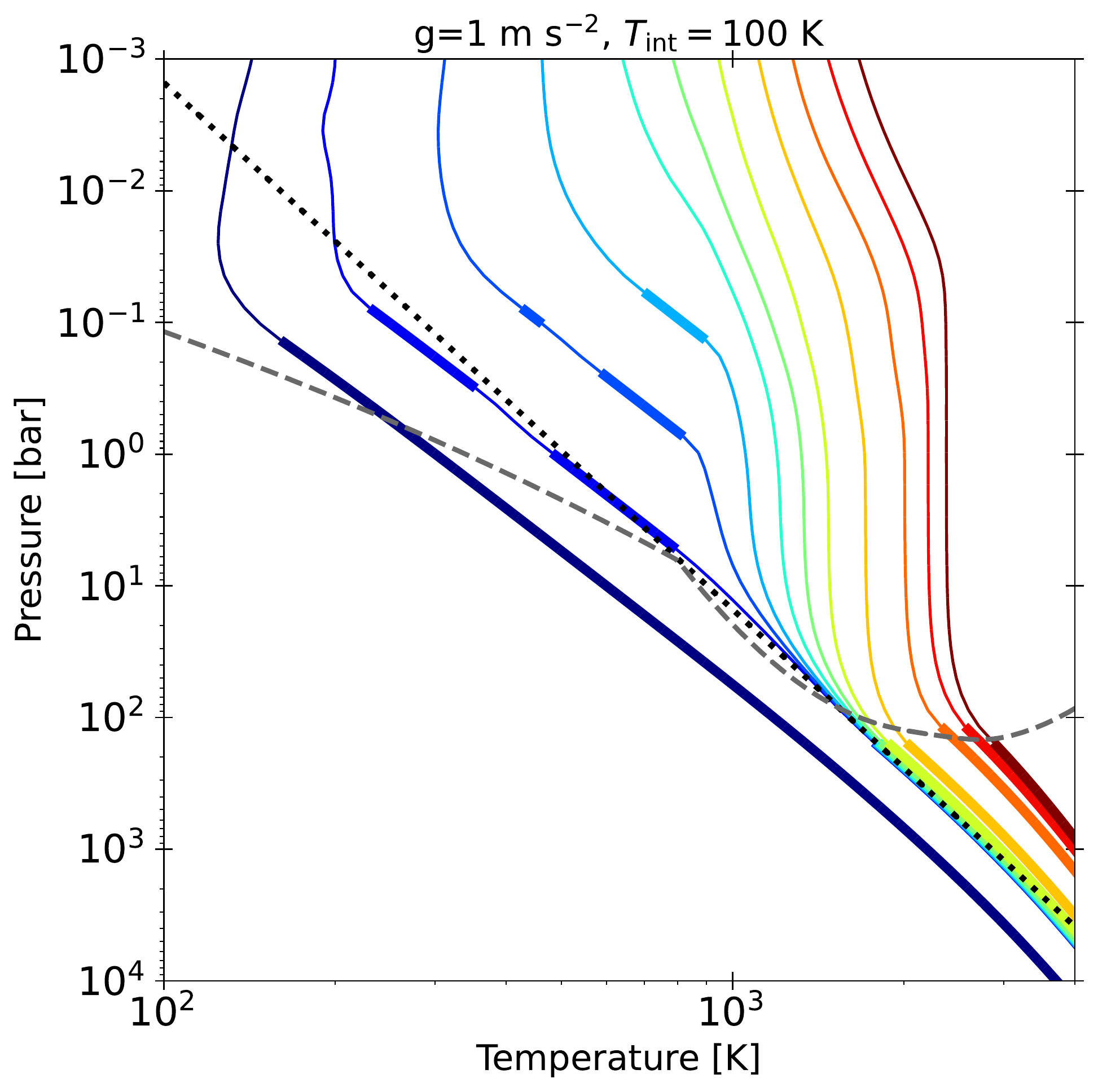}
\includegraphics[clip, width=0.33\hsize]{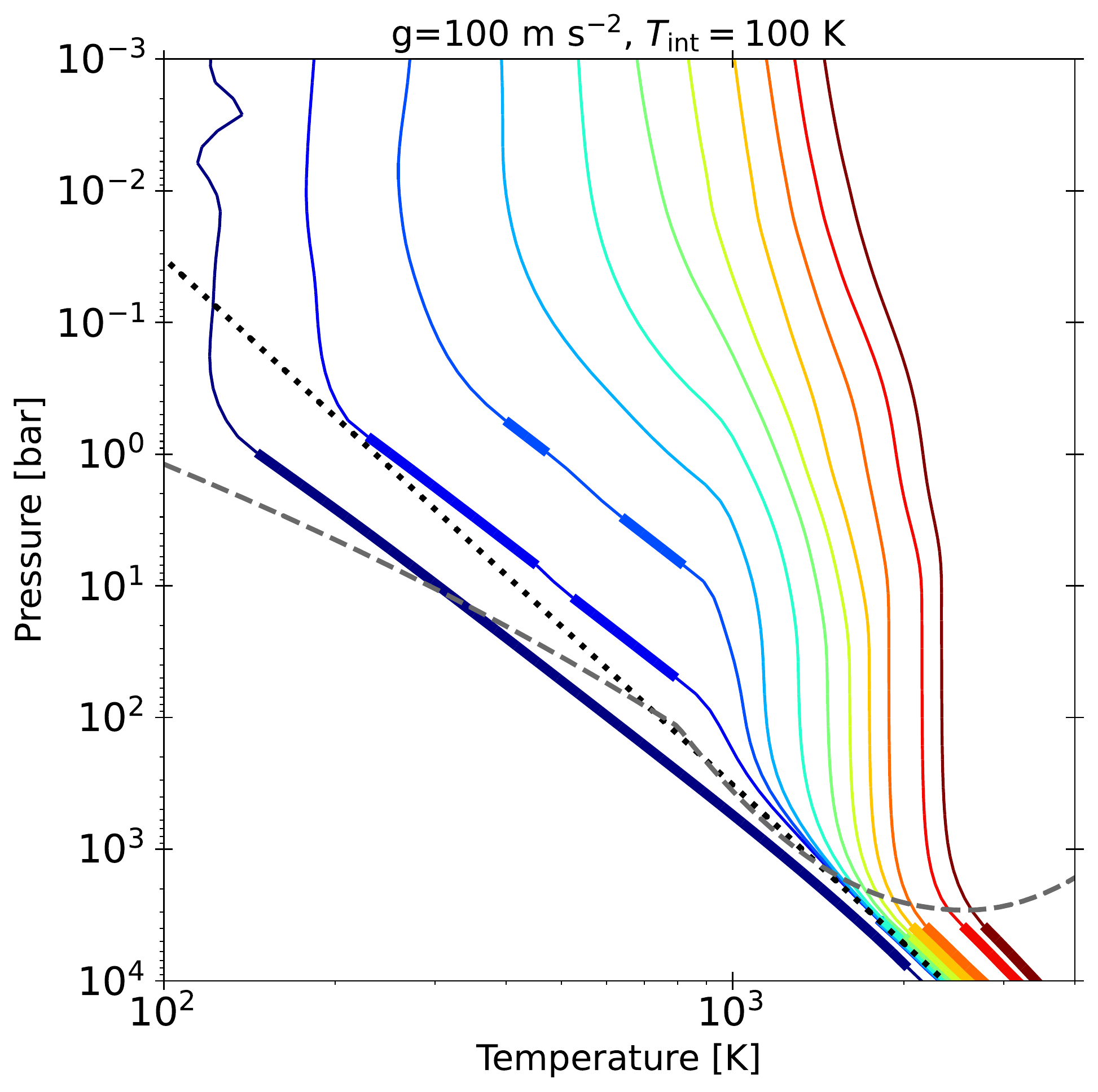}
\includegraphics[clip,width=0.33\hsize]{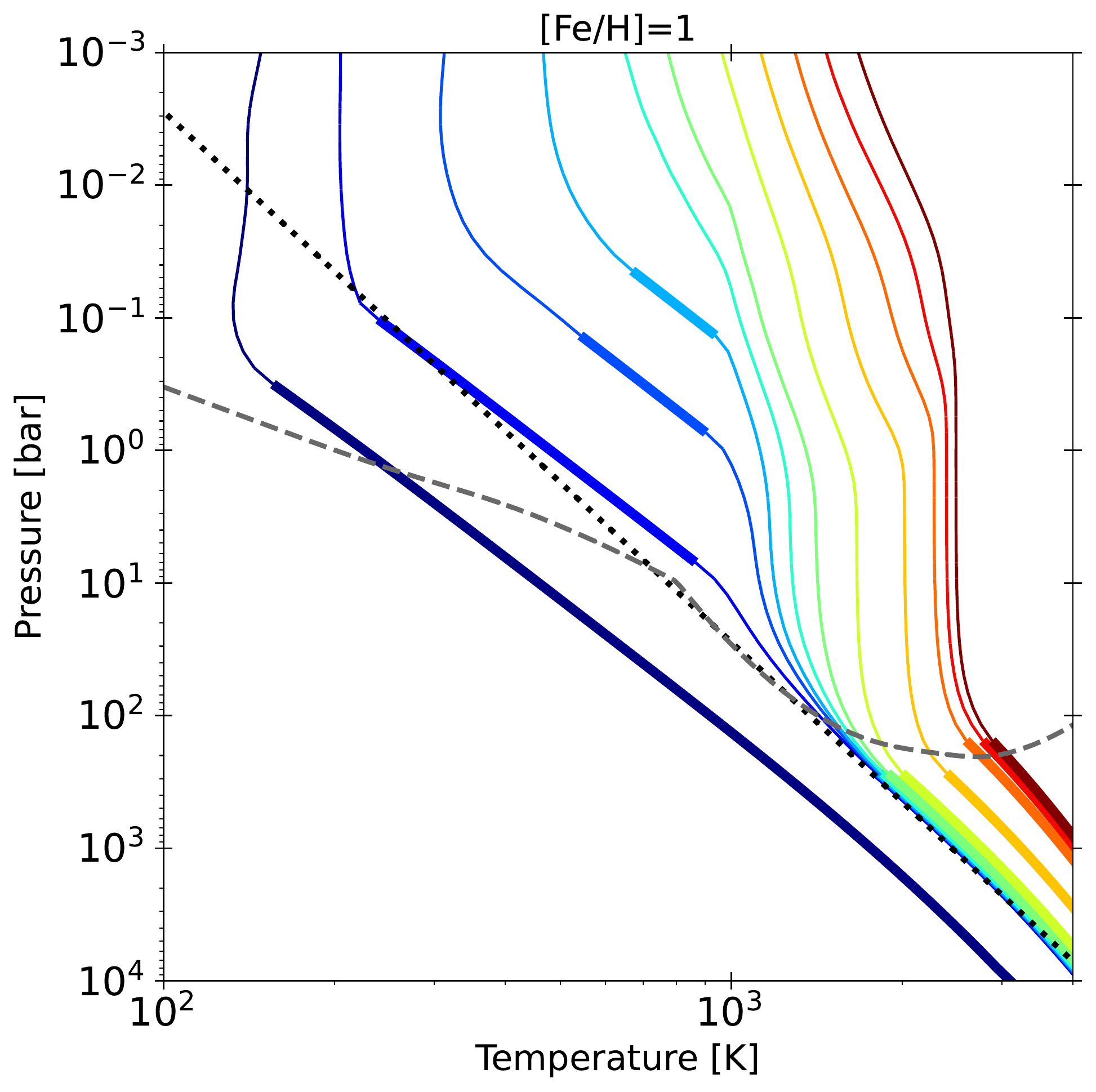}
\includegraphics[clip,width=0.33\hsize]{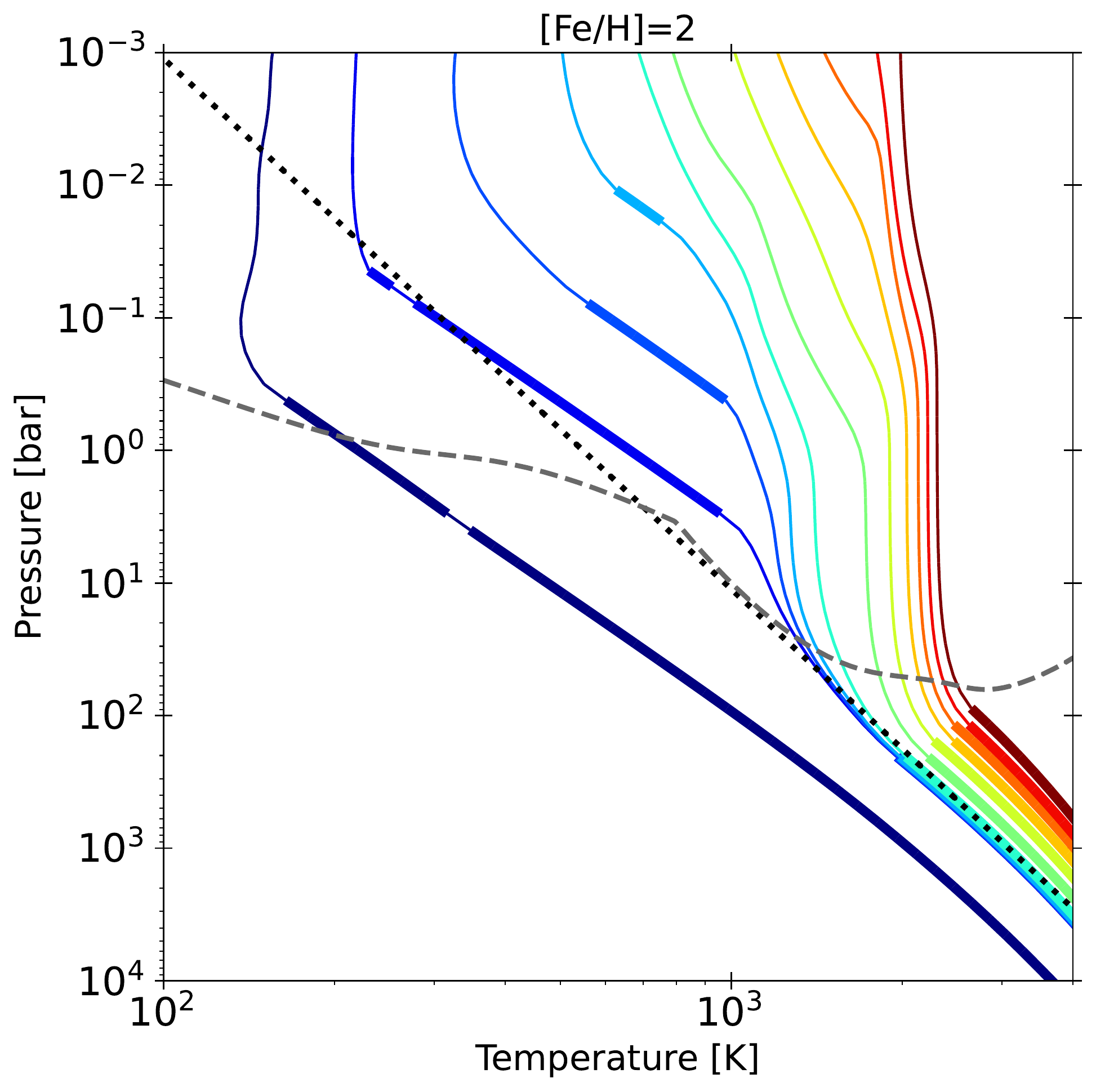}
\includegraphics[clip,width=0.33\hsize]{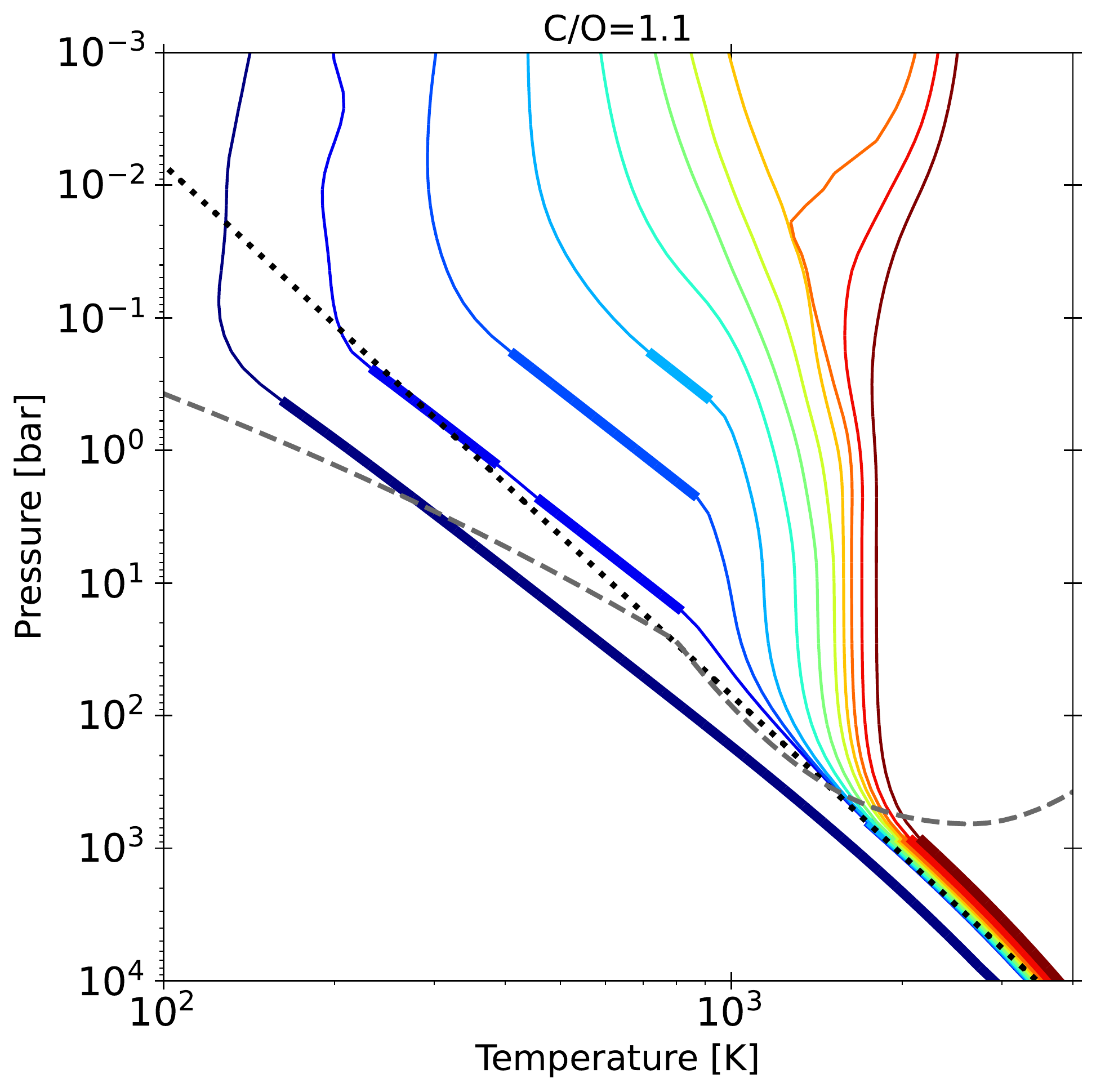}
\caption{%Same as Figure \ref{fig:PT_summary}, but for different planetary intrinsic temperature and surface gravity.
\emph{P--T} profiles for a variety of planetary surface gravities, intrinsic temperatures, atmospheric metallicities, and C/O ratios.  All models are 1D radiative-convective equilibrium. The thicker lines denote convective regions, and thin lines show the radiative regions. The gray dashed line plots the RCB pressure estimated from Equation \eqref{eq:P_RCB1} with the Rosseland mean opacity of \citet{Freedman+14}, which reasonably explains the deep innermost RCB found in numerical results. The black dotted line shows our semi-analytical fit of the deep adiabatic \emph{P--T} profile (Equation \ref{eq:Pad_fit}), which the \emph{P--T} profiles converge on, for $T_{\rm eq}\sim250$--$1200~{\rm K}$.}
\label{fig:PT_summary}
\end{figure*}
%%%%%%%%%%%%%%%%%%%%%%%%%%%%%

The radiative zero solution does not necessarily apply for atmospheric structures, as convection sets in to force the temperature gradient to the adiabatic temperature gradient $\nabla_{\rm ad}$.
In the convective region, from the definition of the adiabatic gradient $(d\ln{T}/d\ln{P})=\nabla_{\rm ad}$, the temperature structure follows
\begin{equation}
    T = T_{\rm rcb}\left( \frac{P}{P_{\rm rcb}}\right)^{\nabla_{\rm ad}},
\end{equation}
where $P_{\rm rcb}$ and $T_{\rm rcb}$ are pressure and temperature of radiative-convective boundary (RCB), and we assume a constant adiabatic gradient for the sake of simplicity.
Inserting Equation \eqref{eq:rad_zero} into this equation with $P=P_{\rm rcb}\gg P_{\rm 0}$, the deep adiabatic temperature can be expressed by
\begin{equation}\label{eq:T_ad}
    T = \left[ T_{\rm 0}+ \left(\frac{3\kappa_{\rm 0}T_{\rm int}^4P_{\rm rcb}^{1+\alpha}(4-\beta)}{16g(1+\alpha)} \right)^{1/(4-\beta)}\right]\left( \frac{P}{P_{\rm rcb}}\right)^{\nabla_{\rm ad}},
\end{equation}
This equation could strongly depend on the upper boundary condition if the first term in the prefactor (i.e., $T_{\rm 0}$) dominates over the second term.
In other words, the deep adiabatic profile does depend on the upper boundary condition if the atmospheric \emph{P--T} profile meets the RCB before it converges to the radiative-zero solution.

Based on the preceding argument, we suggest that planets have the common deep adiabatic profile regardless of stellar insolation if the \emph{P--T} profile \rev{converges to the radiative zero solution above the RCB pressure level. }%}meets the RCB after it converges to the radiative zero solution. 
Equating Equation \eqref{eq:nabla_rad2} and $\nabla_{\rm ad}$, the RCB pressure is given by
\begin{equation}\label{eq:P_RCB1}
    P_{\rm rcb}=\left(\frac{16gT_{\rm rcb}^{4-\beta}}{3 \kappa_{\rm 0}T_{\rm int}^4} \nabla_{\rm ad}\right)^{1/(1+\alpha)}.
\end{equation}
\rev{Inserting Equation \eqref{eq:P_RCB1} into \eqref{eq:rad_zero}, we obtain the relation between the RCB temperature and upper boundary temperature $T_{\rm 0}$ as
\begin{equation}\label{eq:T0Trcb}
\frac{T_{\rm 0}}{T_{\rm rcb}}\approx 1-\phi^{1/(4-\beta)},
\end{equation}
where we have approximated $P_{\rm 0}=0$, as $P_{\rm rcb}\gg P_{\rm 0}$. We have introduced a dimensionless parameter defined as
\begin{equation}\label{eq:phi}
    \phi \equiv \frac{4-\beta}{1+\alpha}\nabla_{\rm ad}.
\end{equation}
The $\phi$ parameter is equivalent to the ratio of adiabatic temperature gradient to the radiative temperature gradient in the limit of deep atmospheres.
Equation \eqref{eq:T0Trcb} is invalid at $\phi>1$ because one cannot define an RCB in the atmosphere with $\phi>1$, where convection does not occur (see Appendix \ref{sec:appendix}).
%The $\phi$ parameter quantifies how much the radiative-zero solution ($T\propto P^{(1+\alpha)/(4-\beta)}=P^{\nabla_{\rm ad}/\phi}$) resembles the adiabatic profile ($T\propto P^{\nabla_{\rm ad}}$).
}
Meanwhile, solving the equality of the first and second terms in the prefactor of Equation \eqref{eq:T_ad} with respect to $P_{\rm rcb}$, we can evaluate a threshold RCB pressure above which the thermal structure converges to the radiative zero solution before it meets the RCB, as
\begin{equation}\label{eq:RCB_cri}
    P_{\rm thr}=\left[ \frac{16gT_{\rm 0}^{4-\beta}(1+\alpha)}{3\kappa_{\rm 0}T_{\rm int}^4(4-\beta)} \right]^{1/(1+\alpha)}.
\end{equation}
Taking the ratio of Equation \eqref{eq:P_RCB1} to \eqref{eq:RCB_cri} \rev{with \eqref{eq:T0Trcb}, }%with an approximation of $T_{\rm rcb}\approx T_{\rm 0}$, 
we achieve the diagnostic metric, given by
\rev{
\begin{eqnarray}\label{eq:metric_adiabat}
    \nonumber
    \frac{P_{\rm rcb}}{P_{\rm thr}}&=&\left( \frac{4-\beta}{1+\alpha}\nabla_{\rm ad}\right)^{1/(1+\alpha)}\left(\frac{T_{\rm rcb}}{T_{\rm 0}}\right)^{(4-\beta)/(1+\alpha)}\\
    &=&\left[ \phi^{1/(\beta-4)}-1\right]^{(\beta-4)/(\alpha+1)}
\end{eqnarray}}
If $P_{\rm rcb}/P_{\rm thr}>1$, the \emph{P--T} profile converges to the radiative zero solution before it meets the RCB, resulting in a deep adiabatic profile being independent of upper boundary condition, i.e., level of stellar insolation.
Interestingly, Equation \eqref{eq:metric_adiabat} indicates that whether or not stellar insolation affects the deep profile only depends on the adiabatic gradient and atmospheric opacity law.
%For example, we assume the constant opacity ($\alpha=\beta=0$) and ideal equation of state with adiabatic index of $7/5$ for diatomic gasses ($\nabla=2/7$), 

Next we investigate Equation \eqref{eq:metric_adiabat} using the opacity of solar composition gas, from \citet{Freedman+14}.
We numerically compute $\alpha$ and $\beta$ from the analytical fit of the Rosseland-mean opacity obtained by \citet{Freedman+14} as a function of pressure and temperature, as shown in the upper two panels of Figure \ref{fig:dependence}.
While previous studies adopted single values of $\alpha$ and $\beta$ \citep[e.g.,][]{Rogers&Seager10,Owen&Wu17,Ginzburg+18}, these values differ at different pressure and temperature conditions.
We also note that since the complex \citet{Freedman+14} analytic fitting formula changes at 800 K, which is necessary since the opacities change with $T$, this leads to plots of $\beta$ that somewhat exaggerate the sharpness of this change to the opacities.
The adiabatic gradient is taken from the equation of state (EOS) for H/He mixtures \citep{Chabrier+19}, which has updated the widely used SCvH EOS \citep{Saumon+95}.
%For the sake of simplicity, we adopt the adiabatic gradient of an analytic fit to \citet{Saumon+95} obtained by \citet{Parmentier+15}:
%\begin{equation}\label{eq:nabla_ad}
%    \nabla_{\rm ad}\approx 0.32-0.1\left( \frac{T}{3000~{\rm K}}\right).
%\end{equation}
The lower left panel of Figure \ref{fig:dependence} shows $P_{\rm rcb}/P_{\rm thr}$ as a function of pressure and temperature. 
%We find that the RCB pressure is comparable or even higher than the threshold pressure in the temperature range of $\sim800$--$1500~{\rm K}$.
We find that the RCB pressure is \rev{much higher} than the threshold pressure in the temperature range of $\sim800$--$1400~{\rm K}$.
\rev{Actually, the $\phi$ parameter takes a value of $\phi\ga1$ in that temperature range, which prohibits the transition from radiative to convective atmospheres.}
This indicates that the \emph{P--T} profiles tend to converge to the radiative zero solution before reaching the RCB in that temperature range.
In Figure \ref{fig:PT_example}, \emph{P--T} profiles indeed converge to the same deep adiabatic profile when the temperature at the second (deeper) nearly isothermal region at $P\sim 10$--$100~{\rm bar}$ falls into $\sim800$--$1500~{\rm K}$, consistent with the phase space of \rev{$P_{\rm rcb}/P_{\rm thr}\gg 1$} in Figure \ref{fig:dependence}.
\citet{Fortney+07} and \citet{Fortney+20} also obtained the \emph{P--T} profiles converging to same deep adiabatic line when the same condition applies.

We note that the deep adiabatic profile might diverge even if the \emph{P--T} profile converges to the radiative zero solution.
This is because the temperature structure obeys $T\propto P^{(1+\alpha)/(4-\beta)}$ in the radiative zero solution (Equation \ref{eq:rad_zero}), where different power-law index of $(1+\alpha)/(4-\beta)$ yields different temperature structure lines. 
As shown in the bottom right panel of Figure \ref{fig:dependence}, the index is roughly constant, $(1+\alpha)/(4-\beta)\sim 0.25$, in the temperature range of $\sim800$--$1300~{\rm K}$.
Thus, it might be expected that the \emph{P--T} profiles tend to converge to the radiative-zero solution of $T\propto P^{0.25}$ in this temperature range where Equation \eqref{eq:metric_adiabat} predicts $P_{\rm rcb}\ga P_{\rm thr}$.
%Thus, it might be expected that the \emph{P--T} profiles tend to converge to the same \emph{P--T} profile of $T\propto P^{0.25}$ in that temperature range.
%We may estimate the equilibrium temperature range at which the PT profiles converge to the same deep adiabatic line as follows.
%Under the approximation of constant atmospheric thermal opacity and visible-to-thermal opacity ratio, \citet{Guillot10} derived an analytical solution of radiative atmosphere given by
%\begin{equation}
    %T^4 = \frac{3T_{\rm int}^4}{4}\left[ \frac{2}{3}+\tau\right]+\frac{3T_{\rm irr}^4}{4}f\left[ \frac{2}{3}+\frac{1}{\gamma \sqrt{3}}+\left(\frac{\gamma}{\sqrt{3}}-\frac{1}{\gamma \sqrt{3}} \right)e^{-\gamma \tau \sqrt{3}}\right],
%\end{equation}
%where $T_{\rm irr}=\sqrt{2}T_{\rm eq}$ is the irradiation temperature, $f$ is the heat redistribution factor, $\tau$ is the thermal optical depth, and $\gamma$ is the visible-to-thermal opacity ratio.
%Note that this equation asymptotically converges to the same PT profile of $T^4=(3/4)T_{\rm irr}^4\tau$ regardless of $T_{\rm irr}$ in the limit of $\tau \rightarrow \infty$.
%Setting $T_{\rm int}=0$ and $\tau \rightarrow \infty$, we can also estimate the temperature at the second isothermal region as (see Section 2.5 of \citealt{Guillot10})
%\begin{equation}
%    \frac{T}{T_{\rm irr}}=\left[ \frac{3f}{4}\left( \frac{1}{\gamma \sqrt{3}} + \frac{2}{3}\right) \right]^{1/4}
%\end{equation}

%%%%%%%%%%%%%%%%%%%%%%%%%%%%%%%%%
\subsection{Numerical Exploration of P--T profiles}\label{sec:RCE_result}
To further study the thermal structure of deep atmospheres, we explore the PT profiles at wide range of planetary properties using EGP, a version of the 1D radiative-convective equilibrium model of \citet{McKay+89} and \citet{Marley&McKay99} \footnote{\rev{A Python version of the adopted model has now been made publicly available \citep{Mukherjee+22a}.}}.
%to verify the concept of the universal deep adiabat. 
%We compute the PT profiles using a radiative-convective equilibrium model of \citet{McKay+89} that
The model solves for radiative-convective equilibrium in a plane-parallel atmosphere based on the algorithm of \citet{Toon+89} with thermochemical equilibrium accounting for rain-out effects \citep{Fegley&Lodders94,Lodders&Fegley02,Visscher+06,Visscher&Lodders10}. 
The model implements non-gray atmospheric opacity with the correlated-k approximation, where we adopt correlated k-coefficients datasets calculated for 1060 pressure-temperature grid points (\citealt{Lupu+21}, see references therein for the details of opacity sources).%where the details of line opacity list can be found in \citet{Freedman+14,Lupu+14}.
We note that our calculations omit TiO/VO opacity, except for the C/O$=1.1$ models.
In convective layers, the model switches to use the adiabatic temperature gradient extracted from the equation of state for H/He mixture with $Y=0.292$ in \citet{Chabrier+19}.
The model has been extensively applied for solar system objects \citep{McKay+89,Marley&McKay99,Fortney+11}, exoplanets \citep{Fortney+05,Fortney+07,Fortney+08,Fortney+20,Morley+13,Morley+15,Morley+17,Thorngren+19,Gao+20,Mayorga+21}, and brown dwarfs \citep{Marley+96,Marley+21,Saumon&Marley08,Morley+12,Morley+14,Robinson&Marley14,Tang+21,Karalidi+21,Mukherjee+22b}.
We refer readers to \citet{Marley&Robinson15} and \citet{Marley+21} for further details of the radiative-convective equilibrium model.
%A slightly high value of $Y$ mimics the effect of metal \citep{Chabrier&Debras21}.
%\subsection{Method}
%We use a publicly available disequilibrium chemistry code, VULCAN \citep{Tsai+17}, to simulate the atmospheric chemical abundances for the range of relevant parameters, namely equilibrium temperature, intrinsic temperature, surface gravity, and atmospheric metallicity.

Figure \ref{fig:PT_summary} shows \emph{P--T} profiles for various values of the planetary equilibrium temperature, surface gravity, intrinsic temperature, atmospheric metallicity, and C/O ratio.
As found in previous studies, cooler planets ($T_{\rm eq}\la1000~{\rm K}$) tend to have steeper temperature gradients, which yields hotter middle atmospheres ($P\sim0.1$--$1~{\rm bar}$) as compared to the equilibrium temperature. 
%owing to the relatively efficient infrared cooling as compared to visible heating \citep{Freedman+14}.
The cooler atmospheres ($T_{\rm eq}\la250$--$600~{\rm K}$) also develop detached convective layers at around $P\sim 0.1$--$10~{\rm bar}$.  A convective zone at these pressures is found in non-irradiated models at these $T_{\rm eff}$ values, and is the ``natural'' outcome for these atmospheres, given the high thermal infrared opacities \citep{Marley&Robinson15}.  Such convective zones are not possible in the more highly irradiated objects given the high temperatures at low pressure, which forces a shallower-than-adiabatic temperature gradient throughout much of the atmosphere.  
%For self-luminous objects, \citet{Marley&Robinson15} explained that the detached convective layer is caused by the shift of thermal emission peaks from wavelengths of opacity windows to more opaque wavelengths.
%, which is originated by the shift of thermal emission peaks from wavelengths of opacity windows to more opaque wavelengths \citep{Marley&Robinson15}.

The \emph{P--T} profiles converge to the same deep adiabatic profile in the equilibrium temperature range of $T_{\rm eq}\sim 250$--$1200~{\rm K}$ for a given set of planetary intrinsic temperature, gravity, and atmospheric compositions.
%universal deep adiabat exists at wide range of planetary properties.
For solar metallicity with $g=1$, $10$, and $100~{\rm m^2~s^{-1}}$ and $T_{\rm int}=50$, $100$, and $200~{\rm K}$ (top five panels), planets with the equilibrium temperature of $T_{\rm eq}=250$--$1200~{\rm K}$ have nearly the same \emph{P--T} profiles in deep convective layers.
These atmospheric models have temperatures in the middle atmosphere ($P\sim10$--$100~{\rm bar}$) between $\sim 1000$--$1500~{\rm K}$, which is consistent with the criterion argued in the previous section.
\emph{P--T} profiles also tend to converge to the same adiabatic profile for high metallicity models of $\rm [Fe/H]=+1$ and $+2$, as well as a higher C/O model of $\rm C/O=1.1$.
At very high metallicities, the deep adiabatic profiles starts to deviate from the same deep profile at a lower equilibrium temperature, for instance at $T_{\rm eq}=1000~{\rm K}$ for $\rm [Fe/H]=2$.
This could be attributed to the middle atmosphere temperature being relatively hotter than that for low-metallicity models, which acts to cause the RCB before the temperature structure converges to the radiative zero solution.

\subsection{Semi-analytical Model of the Deep Adiabat}
We now derive a semi-analytical fit to the universal thermal structure of deep atmospheres for $T_{\rm eq}\sim250$--$1200~{\rm K}$.
Since the deep adiabatic profile would be scaled by the RCB, we infer the parameter dependence of $P\propto (\kappa_{\rm 0}g)^{1/(1+\alpha)}T_{\rm int}^{-4/(1+\alpha)}T^{(4-\beta)/(1+\alpha)}$ from Equation \eqref{eq:P_RCB1}.
According to Figure \ref{fig:dependence}, the opacity law approximately follows $\alpha\sim0.5$ and $(4-\beta)/(1+\alpha)\sim4$ in the temperature range of interest.
In addition, the reference opacity $\kappa_{\rm 0}$ depends on the metallicity.
We assume the metallicity dependence of $\kappa_{\rm 0} \propto {10}^{c{\rm [Fe/H]}}$, where $c$ is a fitting constant \footnote{\rev{This dependence is motivated by the analytic Rosseland mean opacity of \citet{Freedman+14}, who also assumed the opacity proportional to $10^{c{\rm[Fe/H]}}$. \citet{Freedman+14} considered different $c$ coefficients between high and low pressure limits and also considered a temperature dependence of $c$ at the high pressure limit. Here, We have assumed a constant value of $c$ for the sake of simplicity.}}.
Inserting these values and determining the reference pressure to match numerical results, we achieve the following analytical form of the deep adiabatic \emph{P--T} profile which \emph{P--T} profiles converge at $T_{\rm eq}\sim250$--$1200~{\rm K}$, as
\begin{equation}\label{eq:Pad_fit}
    P\approx 70\times{10}^{-0.4{\rm [Fe/H]}}~{\rm bar}~\left( \frac{T}{1000~{\rm K}}\right)^{4}\left( \frac{g}{10~{\rm m~s^{-2}}}\right)^{2/3}\left( \frac{T_{\rm int}}{100~{\rm K}}\right)^{-8/3},
    %P_{\rm ad}\approx 70~{\rm bar}~\left( \frac{T_{\rm ad}}{1000~{\rm K}}\right)^{4}\left( \frac{g}{10~{\rm m~s^{-2}}}\right)^{1/2}\left( \frac{T_{\rm int}}{100~{\rm K}}\right)^{-2},
\end{equation}
or equivalently
\begin{equation}\label{eq:Pad_fit2}
    T\approx 1090\times{10}^{0.1{\rm [Fe/H]}}~{\rm K}~\left( \frac{P}{100~{\rm bar}}\right)^{1/4}\left( \frac{g}{10~{\rm m~s^{-2}}}\right)^{-1/6}\left( \frac{T_{\rm int}}{100~{\rm K}}\right)^{2/3},
\end{equation}
where we have inserted $c\approx0.6$ to fit the metallicity dependence of the numerical results.
Equation \eqref{eq:Pad_fit} indicates that the deep adiabat becomes hotter at higher intrinsic temperature, metallicity, and lower surface gravity.
\rev{The black dotted lines in Figure \ref{fig:PT_summary} show the analytic deep adiabatic \emph{P--T} profile of Equation \eqref{eq:Pad_fit}.} As seen in the Figure, Equation \eqref{eq:Pad_fit} explains the common deep adiabatic profile for $T_{\rm eq}\sim250$--$1200~{\rm K}$ very well, including its dependence on surface gravity, intrinsic temperature, and atmospheric metallicity. 
Thus, for cool to warm exoplanets with $T_{\rm eq}\sim250$--$1200~{\rm K}$, one can utilize Equation \eqref{eq:Pad_fit} to estimate the thermal structure of the deep atmosphere, such as for estimating the quenched abundance of disequilibrium chemical species.

%%%%%%%%%%%%%%%%%%%%%%%%%%%%%%%%%%%%%%%%%%%%
\section{Exploring the Relationship Between NH$_3$ and Bulk Nitrogen Abundances}\label{sec:N_map}
In the previous section we worked to derive a semi-analytic theory of the deep atmosphere temperature structure as a step towards a semi-analytic understanding of an atmospheres NH$_3$ abundance.  We continue along this path here.  In this section we explore the relationship between observable NH$_3$ and the bulk nitrogen abundance based on semi-analytical arguments.% and photochemical models.

%%%%%%%%%%%%%%%%%%%%%%%%%%%%%
\begin{figure*}[t]
\centering
\includegraphics[clip, width=0.47\hsize]{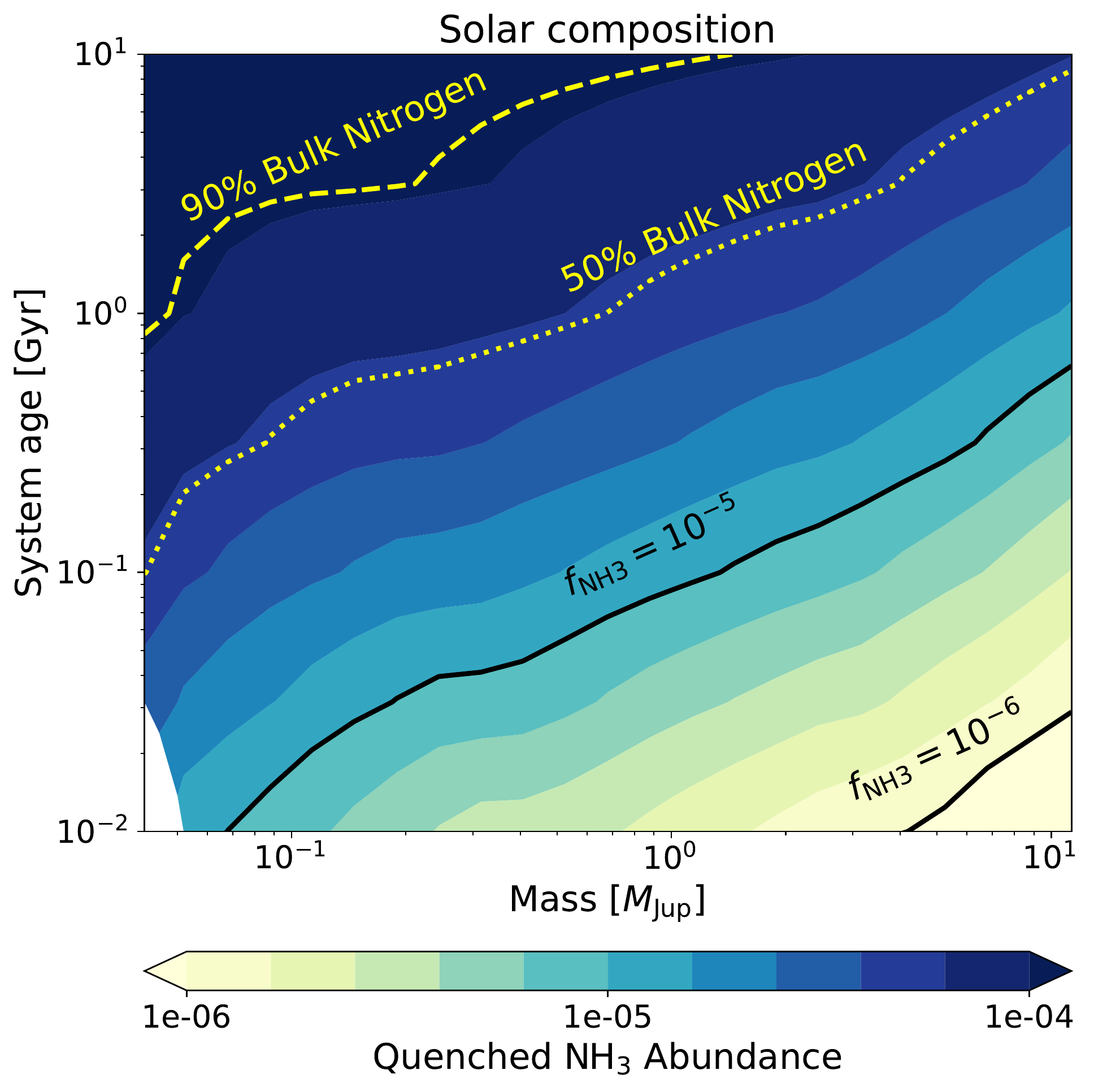}
\includegraphics[clip, width=0.47\hsize]{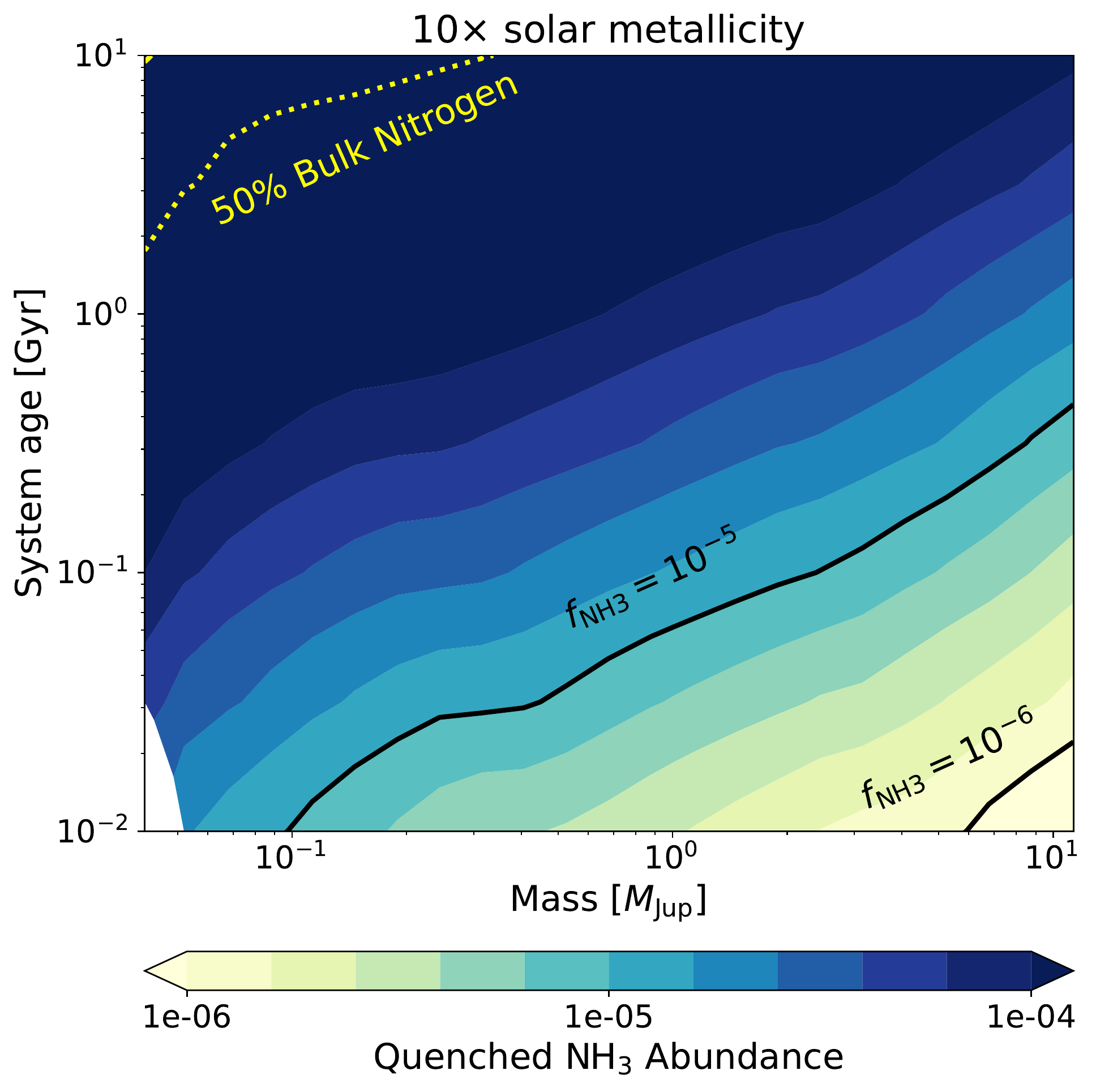}
\includegraphics[clip, width=0.47\hsize]{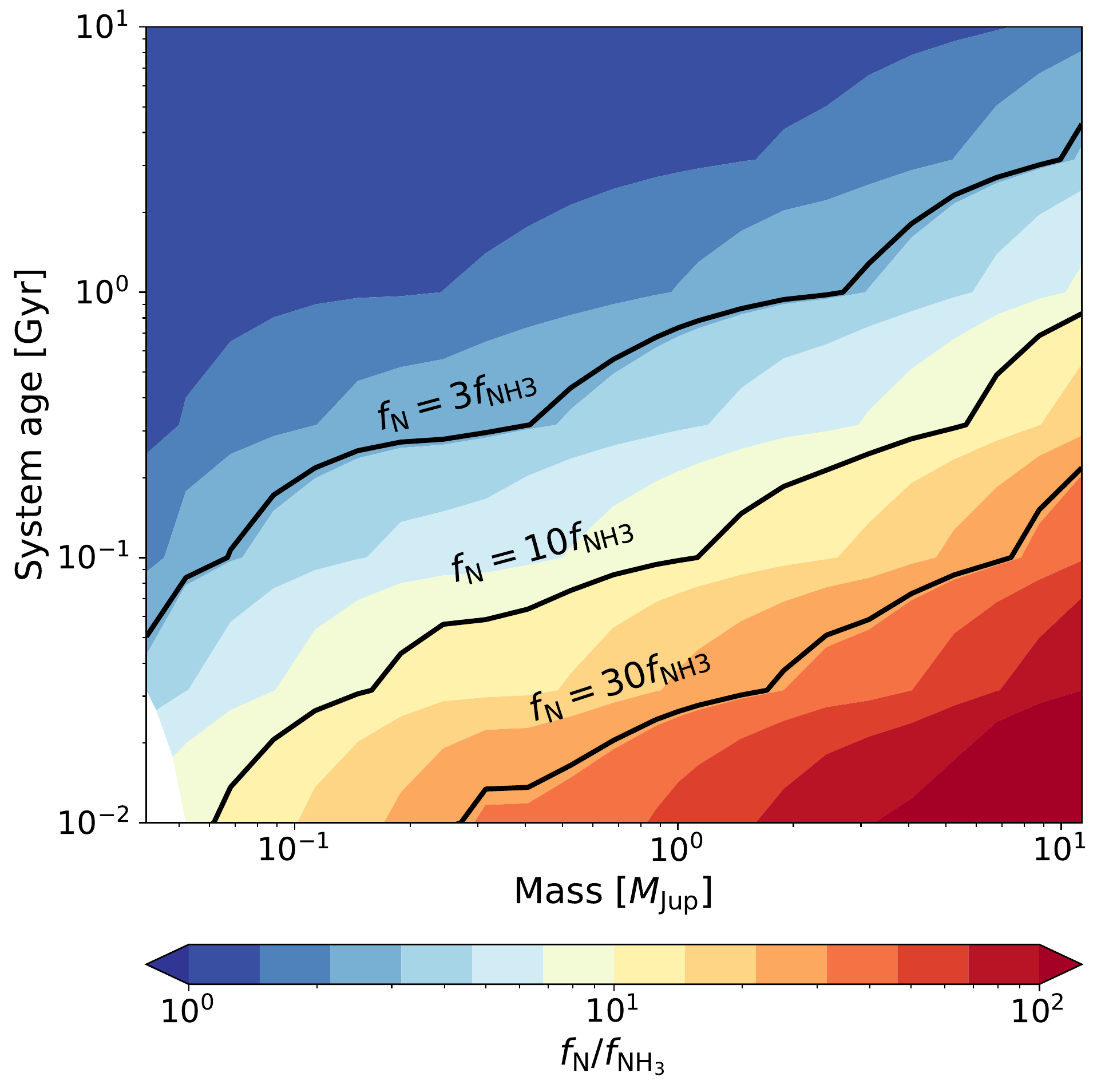}
\includegraphics[clip, width=0.47\hsize]{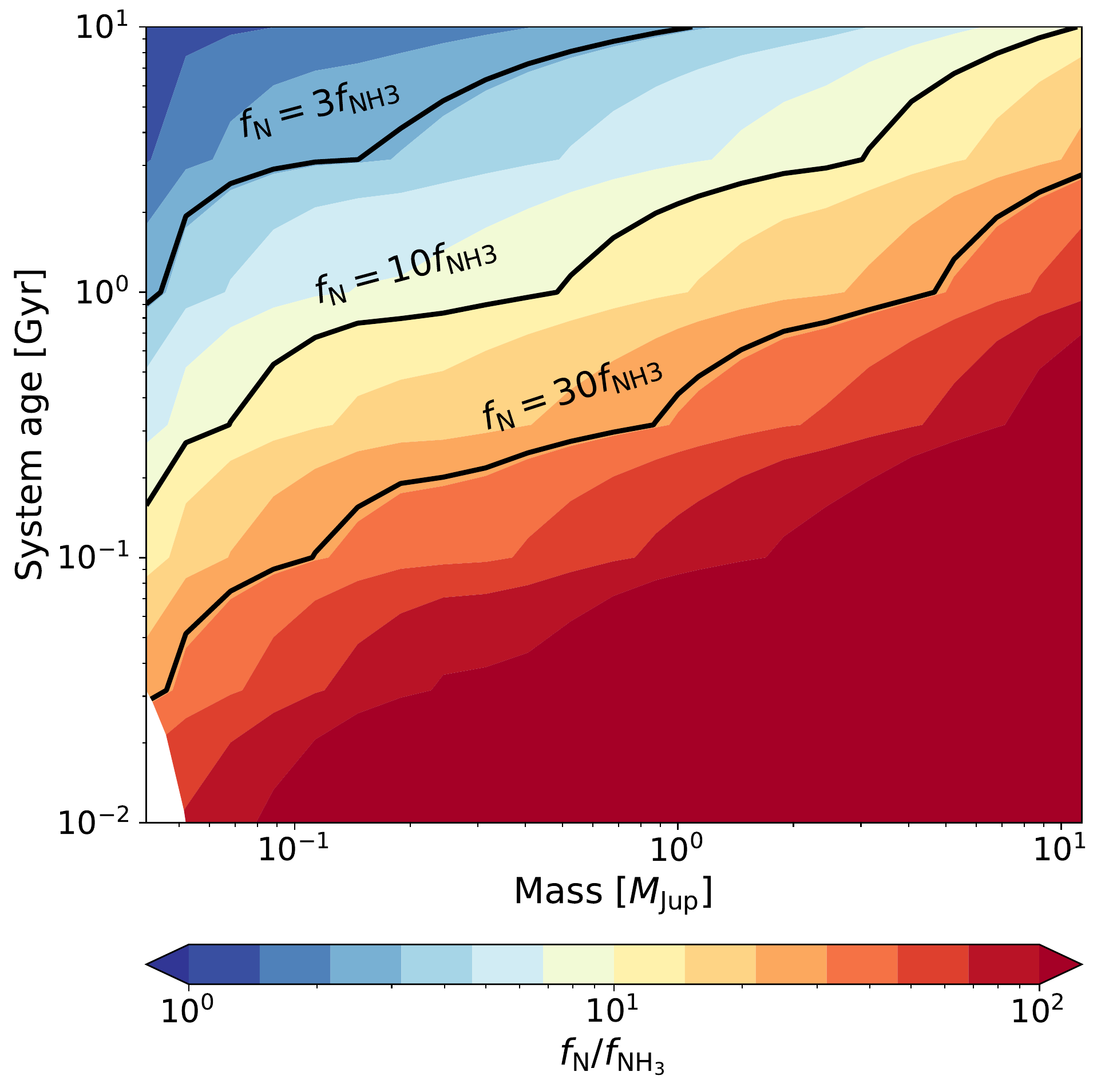}
\caption{The quenched NH$_3$ abundance (top panels, Equation \ref{eq:NH3_analytic}) and the ratio of the bulk nitrogen to the quenched NH$_3$ abundance ratio $f_{\rm N}/f_{\rm NH3}$ (bottom panels, Equation \ref{eq:N_diagnostic}) as a function of a planetary mass and age, \rev{applicable to planets that have the universal deep adiabat ($T_{\rm eq}\sim250$--$1200$ K)}. 
%Quenched NH$_3$ abundance as a function of planetary mass and age predicted by Equation \eqref{eq:NH3_analytic}. 
The black line denote the abundance contours of $f_{\rm NH_3}={10}^{-5}$ and ${10}^{-6}$ for the top panels and the contours of $f_{\rm N}/f_{\rm NH3}=3$, $10$, and $30$ for the bottom panels. \rev{The yellow contours in the upper panels also denote the contours of NH$_3$ abundances corresponding to the 90\% and 50\% of bulk nitrogen budget.} The left and right columns show the results for solar metallicity and $10\times$ solar metallicity atmospheres, respectively, where we have assumed that the N/H ratio is scaled by the metallicity.
%In the higher metallicity atmospheres, the NH$_3$ abundance deviates from the bulk nitrogen abundance in nearly all phase space, except for the very narrow space of $M_{\rm p}<0.1M_{\rm jup}$ and an age of $\ga 10~{\rm Gyr}$.
%Thus, one must be cautious when trying to infer the bulk nitrogen abundances from NH$_3$.
%(Bottom panels) Predicted NH$_3$ fraction as compared to the bulk nitrogen abundance. The black lines denote the contour of $f_{\rm N}/f_{\rm NH_3}=3$, $10$, and $30$. 
%We extract planetary gravity and intrinsic temperature, which is used in Equation \eqref{eq:g_NH3}, as a function of planetary mass and system age from the evolution track of \citet{Fortney+07}.
}
\label{fig:NH3_map}
\end{figure*}
%%%%%%%%%%%%%%%%%%%%%%%%%%%%%

%In this section, we attempt to isolate planetary properties adequate for constraining bulk nitrogen abundance through NH$_3$ observations.
%We first present semi-analytical prediction for the relation between observable NH$_3$ abundance and bulk nitrogen abundance.
%Then, we test our prediction using a disequilibrium photochemical model.

\subsection{Semi-analytical predictions}\label{seq:N_analytic}
We first estimate how the vertically quenched NH$_3$ abundance relates to the bulk nitrogen abundance based on the semi-analytical argument established in previous sections.
Since exoplanets with $T_{\rm eq}\sim250$--$1200~{\rm K}$ have nearly the same deep adiabatic profile (Section \ref{sec:deep_adiabat}) and $\rm NH_3/N_2$ ratio is nearly constant along the deep adiabat (Section \ref{sec:nitrogen_chem}), the quenched NH$_3$ abundance would be nearly independent of the equilibrium temperature, as \rev{previously} demonstrated in \citet{Fortney+20}.
For cool to warm exoplanets with $T_{\rm eq}\sim250$--$1200~{\rm K}$, solving Equation \eqref{eq:f_N1} with respect to $f_{\rm NH3}$, we predict the quenched NH$_3$ abundance of
\begin{equation}\label{eq:NH3_analytic}
    \frac{f_{\rm NH3}}{f_{\rm N}}=\frac{\sqrt{1+8\mathcal{K}^{-1}}-1}{4}\mathcal{K},
\end{equation}
where 
\begin{eqnarray}\label{eq:K}
    \mathcal{K}&=& P^2 f_{\rm H2}^3f_{\rm N}^{-1} Ae^{B/T}\\
    \nonumber
    &\approx& 3.46\times{10}^{-0.8{\rm [Fe/H]}}\left( \frac{f_{\rm N}}{10^{-4}}\right)^{-1}\left( \frac{g}{10~{\rm m~s^{-2}}}\right)^{4/3}\left( \frac{T_{\rm int}}{100~{\rm K}}\right)^{-16/3},
\end{eqnarray}
where we have inserted the semi-analytic deep adiabat (Equation \ref{eq:Pad_fit}) with \rev{$T=2000~{\rm K}$, where the temperature is chosen arbitrary as the equilibrium NH$_3$ abundance is approximately constant along the deep adiabat.}.
Qualitatively speaking, low intrinsic temperature, low atmospheric metallicity, and high surface gravity lead to colder deep atmospheres, which corresponds to large $\mathcal{K}$.
Thus, Equation \eqref{eq:NH3_analytic} yields NH$_3$ rich deep atmospheres of $f_{\rm NH_{\rm 3}}=f_{\rm N}$ in the limit of high $\mathcal{K}$ and vice versa for low $\mathcal{K}$.
Meanwhile, substitution of Equation \eqref{eq:Pad_fit} into \eqref{eq:f_N1} with $T_{\rm }=2000~{\rm K}$ yields
\begin{eqnarray}\label{eq:N_diagnostic}
    %f_{\rm N}&\approx&f2_{\rm NH_3}\left[ 1 + \frac{2f_{\rm NH_3}e^{-B/(2000~{\rm K})}}{Af_{\rm H_2}^2\times 1254400\times {10}^{-0.8{\rm [Fe/H]}}~{\rm bar^2}} \left( \frac{g}{10~{\rm m~s^{-2}}}\right)^{-4/3}\left( \frac{T_{\rm int}}{100~{\rm K}}\right)^{16/3}\right]\\
    %\nonumber
    \frac{f_{\rm N}}{f_{\rm NH_3}}&\approx&1 + 0.58\times{10}^{0.8{\rm [Fe/H]}}\left( \frac{f_{\rm NH_3}}{10^{-4}}\right)\left( \frac{g}{10~{\rm m~s^{-2}}}\right)^{-4/3}\left( \frac{T_{\rm int}}{100~{\rm K}}\right)^{16/3}.
\end{eqnarray}
Equation \eqref{eq:N_diagnostic} enables us to constrain the bulk nitrogen abundance for a given quenched NH$_3$ abundance, atmospheric metallicity, surface gravity, and intrinsic temperature.
The former three values could be constrained by observations, while the intrinsic temperature could be constrained either by thermal evolution models \citep[e.g.,][]{Guillot+96,Burrows+97,Guillot&Showman02,Baraffe+03,Fortney+07,Mordasini+12,Valencia+13,Lopez&Fortney14,Kurosaki+14,Kurokawa&Nakamoto14,Vazan+15,Thorngren+16,Chen&Rogers16,Kubyshkina+20} and/or emission spectroscopy in the limit of very high $T_{\rm int}$ \citep{Morley+17}.

Focusing on cool to warm exoplanets, we here predict the quenched NH$_3$ abundance and its fraction to the bulk nitrogen, $f_{\rm NH_3}/f_{\rm N}$, over a wide range of planetary mass and age.
We combine Equations \eqref{eq:NH3_analytic} and \eqref{eq:N_diagnostic} with the thermal evolution tracks of \citet{Fortney+07} to predict $g$ and $T_{\rm int}$ for given planetary masses and ages, where we adopted the evolution track for the core mass of $10~M_{\rm \oplus}$ and semi-major axis of $0.1~{\rm AU}$ \footnote{Grids of the evolution tracks are available at \url{https://www.ucolick.org/~jfortney/models.htm}}.
Figure \ref{fig:NH3_map} shows the predicted NH$_3$ abundance and its fraction of the bulk nitrogen abundance.
The quenched NH$_3$ abundance is in general higher at lower mass and older planets, as these planets have cooler interiors and deep atmospheres that allow an NH$_3$-rich deep atmosphere.
In many cases, the quenched NH$_3$ abundance exceeds what is a potentially detectable mixing ratio \footnote{\rev{We note that the actual detectable abundance would depend on a number of other factors, such as wavelength range, spectral resolution, chemical species of interest, and abundances of other chemical species.}} of $\ga {10}^{-6}$ (see \citealt{Fortney+20} for the discussion on the threshold), except for super-Jupiter mass planets at very young ages of $\la 0.01~{\rm Myr}$.
In terms of the NH$_3$ fraction to the total nitrogen, for solar metallicity atmospheres (left column of Figure \ref{fig:NH3_map}), the quenched NH$_3$ abundance is almost identical to the bulk nitrogen abundance if the planet has a sub-Jupiter mass ($\la 1M_{\rm j}$) and old ages ($\ga 1~{\rm Gyr}$). 
%Thus, these planets would be good targets to constrain bulk nitrogen abundance from the observed NH$_3$ abundance.
For more massive and younger planets, the quenched NH$_3$ abundance starts to deviate from the bulk nitrogen abundance.
For example, the NH$_3$ abundance is approximately an order of magnitude lower than the bulk nitrogen abundance in Jupiter-mass planets at $0.1~{\rm Gyr}$.
Thus, for massive and young planets, the observed NH$_3$ abundance would only constrains the lower limit of the bulk nitrogen abundance. 
%\rev{This result supports \citet{Fortney+20} who found that the quenched NH$_3$ abundance is lower at younger planets and than the bulk nitrogen abundance in most of their parameter space.}
%, as suggested by \citet{Fortney+20}.

The discrepancy between the NH$_3$ and bulk nitrogen abundance becomes even larger if the planet has a higher metallicity atmosphere.
The right column of Figure \ref{fig:NH3_map} shows the quenched NH$_3$ abundance and $f_{\rm N}/f_{\rm NH3}$ for 10$\times$ solar metallicity atmospheres.
Interestingly, the expected quenched NH$_3$ abundance is almost comparable to that expected for solar metallicity atmospheres, as N$_2$ is favored for both higher N/H and hotter deep atmospheres due to higher metallicities.
This can also be understood as follows.
Assuming a high metallicity atmosphere with $\mathcal{K}\ll8$ and $f_{\rm N}\propto {10}^{\rm [Fe/H]}$, Equation \eqref{eq:NH3_analytic} approximately yields the NH$_3$ abundance of
\begin{equation}
    f_{\rm NH_{\rm 3}}\approx f_{\rm N}\sqrt{\frac{\mathcal{K}}{2}}\propto 10^{0.1{\rm [Fe/H]}}.
\end{equation}
Thus, the NH$_3$ abundance is insensitive to [Fe/H] for high metallicity atmospheres.
The weak metallicity dependence of $f_{\rm NH3}$ leads to the \emph{fraction} of NH$_3$ to the bulk nitrogen, i.e., $f_{\rm NH3}/f_{\rm N}$, being lower in the higher metallicity atmospheres at a given planetary mass and age. 
\rev{In other words, the fraction of observable nitrogen (i.e., NH$_3$) decreases with an increased atmospheric metallicity because of the conversion of NH$_3$ to N$_2$.}
%meaning the fraction of nitrogen that is not detectable as NH$_3$ increases.
%This explains why NH$_3$ abundance is always lower than the bulk nitrogen abundance in \citet{Fortney+20} which assumed Saturn-mass planets with $10\times$ solar metallicity atmospheres.
Importantly then, it is necessary to assess the overall atmospheric metallicity from other spectral features for correctly inferring the bulk nitrogen abundance from NH$_3$.

%%%%%%%%%%%%%%%%%%%%%%%%%%%%%%%%%%%%%%%%%%%%
\section{Discussion}\label{sec:discussion}

\subsection{\rev{Issues of the Strong Dependence on Metallicity}}\label{sec:metallicity_issue}
%\begin{figure}[t]
%\centering
%\includegraphics[clip, width=\hsize]{Map_NH3_mass-metal.pdf}
%\includegraphics[clip, width=\hsize]{Map_NtoNH3_mass-metal.pdf}
%\caption{
%\rev{The same as Figure \ref{fig:NH3_map}, but here we assume the mass-metallicity relation of Solar System giant planets.}
%}
%\label{fig:NH3_map_mass-metal}
%\end{figure}

\rev{
One of our main findings is the strong metallicity dependence of the ratio of the quenched NH$_3$ abundance to the bulk nitrogen abundance.
Atmospheric metallicity could be constrained by the presence of chemical species sensitive to the metallicity, such as CO$_2$ \citep[e.g.,][]{ERS+22,ERS+22_G395,ERS+22_PRISM} and SO$_2$ \citep[][]{Polman+22,ERS+22_SO2}.
Broad wavelength coverage and the unprecedented precision of JWST may also help to better constrain the metallicity.
Thus, we anticipate that observers can use a planet's observationally constrained metallicity for converting the retrieved NH$_3$ abundance to bulk nitrogen abundance through Equation \eqref{eq:N_diagnostic}.
However, the inference would be further complicated if a planetry atmosphere has strongly non-solar elemental ratios (e.g., C/O$\gg$1), as it may cause a distinct deep adiabatic profile from our semi-analytic \emph{P--T} profile. %, although strongly non-solar ratios of N to other elements (if possible) would complicate this.
}

\rev{
It is difficult to predict whether the quenched NH$_3$ abundance is comparable to the bulk nitrogen abundance before observations.
There a few potential ways to coarsely aid such predictions, however.
Interior structure models could set an upper limit on atmospheric metallicity assuming that the metals in the planetary interior is fully mixed to the atmosphere \citep{Thorngren&Fortney19}.
The estimated upper limit may be used to predict the largest discrepancy between the quenched NH$_3$ and bulk nitrogen abundance.
One might eventually also be able to utilize the relation between planetary mass and atmospheric metallicity suggested from Solar System giant planets \citep[e.g.,][]{Kreidberg+14b,Wakeford+17,Welbanks&Madhusudhan19}. 
%Figure \ref{fig:NH3_map_mass-metal} shows the quenched NH$_3$ abundance and ratio of bulk nitrogen to NH$_3$ abundances, where we assume the mass-metallicity relation given by \citep{Welbanks&Madhusudhan19}
%\begin{equation}
%    {\rm[Fe/H]}=-1.11\log_{\rm 10}{(M_{\rm p}/M_{\rm jup})}+0.38
%\end{equation}
%While the quenched NH$_{\rm 3}$ abundance is higher at smaller planets as similar to Figure \ref{fig:NH3_map}, the discrepancy between NH$_3$ and bulk nitrogen abundances becomes larger at smaller planets.
%The result might indicate a dilemma of NH$_3$: smaller planets are more favored for NH$_3$ detection but more difficult to interpret the retrieved abundance.
However, exoplanets have not shown a clear mass-metallicity relation as of yet \citep{Wakeford&Dalba20,Guillot+22,Edwards+22}. We need better knowledge about the population-level metallicity trend of exoplanetary atmospheres to make a reliable prediction.
}

\subsection{\rev{Caveats}}\label{sec:caveat}
\rev{We have assumed that the NH$_3$ abundance is vertically constant above the quench level. While several studies assumed the same approximation to model the transport-driven disequilibrium chemistry \citep[e.g.,][]{Morley+17,Fortney+20,Mukherjee+22b}, the assumption is not always valid. For example, \citet{Moses+11} showed that NH$_3$ abundance gradually decreases with decreasing pressure above the quench level in hot Jupiter HD189733b.
\citet{Moses+21} also obtained similar NH$_3$ profiles in their pseudo-2D photochemical simulations for many planetary equlibrium temperature.
These vertically nonuniform profile could occur when the eddy diffusion timescale is not sufficiently short as compared to chemical interconversion timescale.
We anticipate that the vertically constant abundance would be reasonable for warm to cool exoplanets where the chemical timescale quickly increases with altitude \citep[see][]{Tsai+18}. However, one should always be encouraged to verify the assumption using kinetic chemical model for a specific planet of interest.
%However, one should keep in mind that our semi-analytical model just provides the first order guess. Detailed investigations with kinetic chemical model is always encouraged to precisely interpret the observed NH$_3$ abundance.
}

\rev{
We have assumed nearly constant equilibrium abundance of NH$_3$ along the deep adiabat.
While the assumption is reasonably valid at hydrogen-dominated substellar atmospheres, the assumption would be no longer valid if an atmosphere has different a primary composition with different adiabatic index.
}

\rev{
We have only considered the transport-induced disequilibrium chemistry, while other physical processes can also affect the NH$_3$ vertical profile. \citet{Molaverdikhani+19} provides an in depth discussion about how the photochemistry and molecular diffusion could cause the discrepancy from vertically constant profile. \citet{Hu21} showed that NH$_3$ tends to be depleted by photodissociation in temperate to cold exoplanets. We investigate the effect of photochemistry and how the observable NH$_3$ abundance relates with bulk nitrogen abundance in our Paper II.
}

\subsection{Relevance for Cold and Directly Imaged Planets}

\rev{Planets that lack strong radiative forcing from their parent star would in some ways be simpler to interpret, from an observational perspective.  First, lacking external forcing, their interiors would cool off somewhat faster, into the NH$_3$ dominated chemical T-P phase space.  While their deep atmospheres would not share the radiative-zero solution, they do have the real added benefit that their intrinsic temperature (the object's effective temperature in this case) and the T-P conditions of their deep atmosphere adiabat can be directly constrained by thermal infrared observations. Moreover, those isolated objects can avoid NH$_3$ depletion by photodissociation, which limits the observability of NH$_3$ for irradiated planets \citep{Hu21}. Nitrogen disequilibrium in the atmospheres of such isolated objects was recently modeled in \citet{Karalidi+21,Mukherjee+22b}. 
}
%Cholla Paper, Sagnick's paper.}

\rev{
For the very coldest planets whether irradiated or not, one needs to further consider relevant condensation physics.
In cold exoplanets where NH$_3$ clouds form, NH$_3$ must be depleted above the NH$_3$ cloud base, like solar system giant planets.
The formation of H$_2$O clouds may also affect NH$_3$ abundances.
Recent microwave observations of Jupiter by JUNO revealed that NH$_3$ abundance is still partly depleted even below the NH$_3$ cloud base \citep{Bolton+17,Li+17}. 
\citet{Guillot+20,Guillot+20b} suggested that such NH$_3$ depletion could be explained by the formation of NH${_3}\cdot$H$_2$O condensate. 
NH$_3$ can also be depleted because of the dissolution into liquid H$_2$O clouds \citep{Hu19}.
Thus, one needs to be cautious in interpreting the NH$_3$ abundances on very cool planets where NH$_3$ and/or H$_2$O clouds potentially form. 
}

\section{Summary}\label{sec:summary}
In this study, we have investigated how observable NH$_3$ abundances relate to bulk nitrogen abundances of exoplanetary atmospheres. %nitrogen chemistry, especially for NH$_3$, mainly on warm exoplanetary atmospheres.
We first identified that irradiated substellar atmospheres follow nearly the same deep adiabatic profile over a wide range of equilibrium temperatures ($T_{\rm eq}\sim250$--$1200~{\rm K}$).
We have derived a semi-analytical model of such a universal deep adiabat (Equation \ref{eq:Pad_fit}) that readily explains the radiative-convective equilibrium model.
Then, we established a semi-analytical model that relates vertically quenched NH$_3$ abundances with the bulk nitrogen abundance of the atmosphere (Equation \ref{eq:N_diagnostic}).
\rev{Based on the semi-analytical model, we predict the relation between the quenched NH$_3$ and bulk nitrogen abundances as a function of planetary mass and age. We verify our semi-analytical model using a photochemical kinetic model in Paper II.}
%Subsequently, we performed a series of photochemical calculations for various values of planetary mass, age, equilibrium temperature, eddy diffusion coefficient, and atmospheric composition to explore the relation between observable NH$_3$ abundance and bulk nitrogen abundance.
Our key findings are summarized as follows:

\begin{enumerate}

\item Irradiated giant planet atmospheres have nearly the same deep adiabatic profile for the equilibrium temperature of $T_{\rm eq}\sim250$--$1200~{\rm K}$ for a given set of planetary gravity and intrinsic temperature.
This is caused by the fact that their atmospheric \emph{P--T} profiles tend to converge to the radiative-zero solution that is independent of the upper boundary conditions before they meet the radiative convective boundary (Section \ref{sec:deep_adiabat}).
Based on the series of radiative-convective equilibrium calculations, we have derived a semi-analytical model of such universal deep adiabats applicable to planets with $T_{\rm eq}\sim250$--$1200~{\rm K}$ (Section \ref{sec:RCE_result}, Equations \ref{eq:Pad_fit} or \ref{eq:Pad_fit2}).

\item We have established a semi-analytical model that relates the vertically quenched NH$_3$ abundance with the bulk nitrogen abundance (Equations \ref{eq:NH3_analytic} and \ref{eq:N_diagnostic}).
%Our semi-analytical model reproduces the vertically quenched NH$_3$ abundance computed by a photochemical kinetic model (Section \ref{sec:N_map}).
\rev{Our model is applicable to warm irradiated giant exoplanets. We are able to readily assess discrepancies between the quenched NH$_3$ and bulk nitrogen abundances.  This aids when attempting to infer the bulk nitrogen abundance from an observed NH$_3$ abundance.}
%Thus, our semi-analytical model would mitigate the discrepancy between the quenched NH$_3$ and bulk nitrogen abundances when inferring the bulk nitrogen abundance from an observed NH$_3$. %provide a first-order guess on the bulk nitrogen abundance from the observed NH$_3$ abundance.%and bulk nitrogen abundances.

\item \rev{At solar composition in a giant planet atmosphere,} the vertically quenched NH$_3$ abundance nearly coincides with the bulk nitrogen abundance \emph{only} when a planet has a sub-Jupiter mass ($\la1~{\rm M_{\rm J}}$) and old age ($\ga 1~{\rm Gyr}$). %Thus, one would be able to reasonably constrain the bulk nitrogen abundance from the retrieved NH$_3$ abundance. 
For planets with super-Jupiter mass and/or age younger than $1~{\rm Gyr}$, in contrast, the quenched NH$_3$ abundance is considerably lower than the bulk nitrogen abundance, as the deep atmosphere is so hot that N$_2$ dominates over NH$_3$.(Section \ref{sec:N_map} and Figure \ref{fig:NH3_map}).

%\item \rev{The quenched NH$_3$ abundance itself is less sensitive to atmospheric metallicity. This is because }

\item \rev{As the atmospheric metallicity increases, while the predicted quenched NH$_3$ mixing ratio remains constant at a given mass and age, the ratio of NH$_3$ to the bulk atmospheric nitrogen abundance decreases significantly. The issue of NH$_3$ only containing a fraction of the bulk nitrogen abundance then occurs across all giant planet phase space, at sub-Jupiter planet masses and old ages.  This ``missing nitrogen'' problem can be corrected with an assessment of the deep atmospheric T-P profile (likely from structure or evolution models) and an overall assessment of atmospheric metallicity from other species, such a C- and O-bearing molecules, or alkali metals.}

\end{enumerate}

%%%%%%%%%%%%%%%%%%%%%%%%%%%%%%%%%%
\section*{Acknowledgements}
We are grateful to anonymous reviewer for their insightful comments that greatly improved the quality of this paper.
We thank Masahiro Ikoma for helpful comments on the thermal structures of deep atmospheres.
We also thank Neel Patel, Xinting Yu, Ben Lew, Eliza Kempton, Yuichi Ito, Yui Kawashima, Shota Notsu, Tatsuya Yoshida, and Akifumi Nakayama for fruitful discussions.
%This project was partly done at the Exoplanet Summer Program administered by the Other Worlds Laboratory at the University of California, Santa Cruz, funded by the Heising-Simons Foundation.
This work benefited from the 2022 Exoplanet Summer Program in the Other Worlds Laboratory (OWL) at the University of California, Santa Cruz, a program funded by the Heising-Simons Foundation.
Most of numerical computations were carried out on PC cluster at Center for Computational Astrophysics, National Astronomical Observatory of Japan.
K.O. was supported by JSPS Overseas Research Fellowship.  J.J.F. is supported by
an award from the Simons Foundation.

\rev{\appendix}
\rev{
\section{Physical meaning of $\phi$ parameter}\label{sec:appendix}
In this appendix, we elaborate on the physical meaning of $\phi$, which is introduced in Section \ref{sec:deep_adiabat} to quantify whether the \emph{P--T} profile converges to the radiative-zero solution above the RCB pressure level.
The radiative temperature gradient (Equation \ref{eq:nabla_rad2}) approaches zero in the limit of $P\rightarrow 0$ and increases with increasing the pressure.
However, the gradient does not increase indefinitely.
It converges to a certain value controlled by the opacity law.
Inserting back the radiative-zero solution (Equation \ref{eq:rad_zero} with $T_{\rm 0}=0$ and $P_{\rm 0}=0$) into \eqref{eq:nabla_rad2}, one can find an asymptotic gradient to which the radiative gradient approaches:
\begin{equation}
    \left( \frac{d\ln T}{d\ln P}\right)_{\rm rad,limit}= \frac{1+\alpha}{4-\beta}.
\end{equation}
Thus, the definition of $\phi$ parameter (Equation \ref{eq:phi}) can be understood as the ratio of the adiabatic gradient $\nabla_{\rm ad}$ to the above asymptotic radiative gradient:
\begin{equation}
    \phi \equiv \frac{4-\beta}{1+\alpha}\nabla_{\rm ad}=\frac{(d\ln T/d\ln P)_{\rm ad}}{(d\ln T/d\ln P)_{\rm rad,limit}}.
\end{equation}
If the asymptotic radiative gradient is much larger than the adiabatic gradient, i.e., $\phi\ll1$, convection would quickly set in before the \emph{P--T} profile converges to the radiative-zero solution with the asymptotic gradient.
By contrast, $\phi>1$ means that the radiative gradient never exceeds the adiabatic gradient even in the limit of deep atmospheres. Thus, convection does not occur for $\phi>1$.
}
%%%%%%%%%%%%%%%%%%%%%%%%%%%%%%%%%%%%%%%%%%%%%%%%%%
\bibliography{references}

\begin{thebibliography}{}
\expandafter\ifx\csname natexlab\endcsname\relax\def\natexlab#1{#1}\fi
\providecommand{\url}[1]{\href{#1}{#1}}
\providecommand{\dodoi}[1]{doi:~\href{http://doi.org/#1}{\nolinkurl{#1}}}
\providecommand{\doeprint}[1]{\href{http://ascl.net/#1}{\nolinkurl{http://ascl.net/#1}}}
\providecommand{\doarXiv}[1]{\href{https://arxiv.org/abs/#1}{\nolinkurl{https://arxiv.org/abs/#1}}}

\bibitem[{{Alderson} {et~al.}(2022){Alderson}, {Wakeford}, {Alam}, {Batalha},
  {Lothringer}, {Adams Redai}, {Barat}, {Brande}, {Damiano}, {Daylan},
  {Espinoza}, {Flagg}, {Goyal}, {Grant}, {Hu}, {Inglis}, {Lee}, {Mikal-Evans},
  {Ramos-Rosado}, {Roy}, {Wallack}, {Batalha}, {Bean}, {Benneke},
  {Berta-Thompson}, {Carter}, {Changeat}, {Col{\'o}n}, {Crossfield},
  {D{\'e}sert}, {Foreman-Mackey}, {Gibson}, {Kreidberg}, {Line},
  {L{\'o}pez-Morales}, {Molaverdikhani}, {Moran}, {Morello}, {Moses},
  {Mukherjee}, {Schlawin}, {Sing}, {Stevenson}, {Taylor}, {Aggarwal}, {Ahrer},
  {Allen}, {Barstow}, {Bell}, {Blecic}, {Casewell}, {Chubb}, {Crouzet},
  {Cubillos}, {Decin}, {Feinstein}, {Fortney}, {Harrington}, {Heng}, {Iro},
  {Kempton}, {Kirk}, {Knutson}, {Krick}, {Leconte}, {Lendl}, {MacDonald},
  {Mancini}, {Mansfield}, {May}, {Mayne}, {Miguel}, {Nikolov}, {Ohno}, {Palle},
  {Parmentier}, {Petit dit de la Roche}, {Piaulet}, {Powell}, {Rackham},
  {Redfield}, {Rogers}, {Rustamkulov}, {Tan}, {Tremblin}, {Tsai}, {Turner}, {de
  Val-Borro}, {Venot}, {Welbanks}, {Wheatley}, \& {Zhang}}]{ERS+22_G395}
{Alderson}, L., {Wakeford}, H.~R., {Alam}, M.~K., {et~al.} 2022, arXiv
  e-prints, arXiv:2211.10488.
\newblock \doarXiv{2211.10488}

\bibitem[{{Ali-Dib} {et~al.}(2014){Ali-Dib}, {Mousis}, {Petit}, \&
  {Lunine}}]{Ali-dib+14}
{Ali-Dib}, M., {Mousis}, O., {Petit}, J.-M., \& {Lunine}, J.~I. 2014, \apj,
  785, 125, \dodoi{10.1088/0004-637X/785/2/125}

\bibitem[{{Asplund} {et~al.}(2021){Asplund}, {Amarsi}, \&
  {Grevesse}}]{Asplund+21}
{Asplund}, M., {Amarsi}, A.~M., \& {Grevesse}, N. 2021, \aap, 653, A141,
  \dodoi{10.1051/0004-6361/202140445}

\bibitem[{{Atreya} {et~al.}(2022){Atreya}, {Crida}, {Guillot}, {Li}, {Lunine},
  {Madhusudhan}, {Mousis}, \& {Wong}}]{Atreya+22}
{Atreya}, S.~K., {Crida}, A., {Guillot}, T., {et~al.} 2022, arXiv e-prints,
  arXiv:2205.06914.
\newblock \doarXiv{2205.06914}

\bibitem[{{Baraffe} {et~al.}(2003){Baraffe}, {Chabrier}, {Barman}, {Allard}, \&
  {Hauschildt}}]{Baraffe+03}
{Baraffe}, I., {Chabrier}, G., {Barman}, T.~S., {Allard}, F., \& {Hauschildt},
  P.~H. 2003, \aap, 402, 701, \dodoi{10.1051/0004-6361:20030252}

\bibitem[{{Bitsch} {et~al.}(2022){Bitsch}, {Schneider}, \&
  {Kreidberg}}]{Bitsch+22}
{Bitsch}, B., {Schneider}, A.~D., \& {Kreidberg}, L. 2022, arXiv e-prints,
  arXiv:2207.06077.
\newblock \doarXiv{2207.06077}

\bibitem[{{Bolton} {et~al.}(2017){Bolton}, {Adriani}, {Adumitroaie}, {Allison},
  {Anderson}, {Atreya}, {Bloxham}, {Brown}, {Connerney}, {DeJong}, {Folkner},
  {Gautier}, {Grassi}, {Gulkis}, {Guillot}, {Hansen}, {Hubbard}, {Iess},
  {Ingersoll}, {Janssen}, {Jorgensen}, {Kaspi}, {Levin}, {Li}, {Lunine},
  {Miguel}, {Mura}, {Orton}, {Owen}, {Ravine}, {Smith}, {Steffes}, {Stone},
  {Stevenson}, {Thorne}, {Waite}, {Durante}, {Ebert}, {Greathouse}, {Hue},
  {Parisi}, {Szalay}, \& {Wilson}}]{Bolton+17}
{Bolton}, S.~J., {Adriani}, A., {Adumitroaie}, V., {et~al.} 2017, Science, 356,
  821, \dodoi{10.1126/science.aal2108}

\bibitem[{{Booth} {et~al.}(2017){Booth}, {Clarke}, {Madhusudhan}, \&
  {Ilee}}]{Booth+17}
{Booth}, R.~A., {Clarke}, C.~J., {Madhusudhan}, N., \& {Ilee}, J.~D. 2017,
  \mnras, 469, 3994, \dodoi{10.1093/mnras/stx1103}

\bibitem[{{Booth} \& {Ilee}(2019)}]{Booth&Ilee19}
{Booth}, R.~A., \& {Ilee}, J.~D. 2019, \mnras, 487, 3998,
  \dodoi{10.1093/mnras/stz1488}

\bibitem[{{Bosman} {et~al.}(2019){Bosman}, {Cridland}, \& {Miguel}}]{Bosman+19}
{Bosman}, A.~D., {Cridland}, A.~J., \& {Miguel}, Y. 2019, \aap, 632, L11,
  \dodoi{10.1051/0004-6361/201936827}

\bibitem[{{Burrows} {et~al.}(1997){Burrows}, {Marley}, {Hubbard}, {Lunine},
  {Guillot}, {Saumon}, {Freedman}, {Sudarsky}, \& {Sharp}}]{Burrows+97}
{Burrows}, A., {Marley}, M., {Hubbard}, W.~B., {et~al.} 1997, \apj, 491, 856,
  \dodoi{10.1086/305002}

\bibitem[{{Chabrier} {et~al.}(2019){Chabrier}, {Mazevet}, \&
  {Soubiran}}]{Chabrier+19}
{Chabrier}, G., {Mazevet}, S., \& {Soubiran}, F. 2019, \apj, 872, 51,
  \dodoi{10.3847/1538-4357/aaf99f}

\bibitem[{{Chachan} {et~al.}(2022){Chachan}, {Knutson}, {Lothringer}, \&
  {Blake}}]{Chachan+22}
{Chachan}, Y., {Knutson}, H.~A., {Lothringer}, J., \& {Blake}, G.~A. 2022,
  arXiv e-prints, arXiv:2211.09080.
\newblock \doarXiv{2211.09080}

\bibitem[{{Chen} \& {Rogers}(2016)}]{Chen&Rogers16}
{Chen}, H., \& {Rogers}, L.~A. 2016, \apj, 831, 180,
  \dodoi{10.3847/0004-637X/831/2/180}

\bibitem[{{Cridland} {et~al.}(2016){Cridland}, {Pudritz}, \&
  {Alessi}}]{Cridland+16}
{Cridland}, A.~J., {Pudritz}, R.~E., \& {Alessi}, M. 2016, \mnras, 461, 3274,
  \dodoi{10.1093/mnras/stw1511}

\bibitem[{{Cridland} {et~al.}(2017){Cridland}, {Pudritz}, {Birnstiel},
  {Cleeves}, \& {Bergin}}]{Cridland+17}
{Cridland}, A.~J., {Pudritz}, R.~E., {Birnstiel}, T., {Cleeves}, L.~I., \&
  {Bergin}, E.~A. 2017, \mnras, 469, 3910, \dodoi{10.1093/mnras/stx1069}

\bibitem[{{Cridland} {et~al.}(2019){Cridland}, {van Dishoeck}, {Alessi}, \&
  {Pudritz}}]{Cridland+19}
{Cridland}, A.~J., {van Dishoeck}, E.~F., {Alessi}, M., \& {Pudritz}, R.~E.
  2019, \aap, 632, A63, \dodoi{10.1051/0004-6361/201936105}

\bibitem[{{Cridland} {et~al.}(2020){Cridland}, {van Dishoeck}, {Alessi}, \&
  {Pudritz}}]{Cridland+20}
---. 2020, \aap, 642, A229, \dodoi{10.1051/0004-6361/202038767}

\bibitem[{{Dash} {et~al.}(2022){Dash}, {Majumdar}, {Willacy}, {Tsai}, {Turner},
  {Rimmer}, {Gudipati}, {Lyra}, \& {Bhardwaj}}]{Dash+22}
{Dash}, S., {Majumdar}, L., {Willacy}, K., {et~al.} 2022, arXiv e-prints,
  arXiv:2204.04103.
\newblock \doarXiv{2204.04103}

\bibitem[{{Drummond} {et~al.}(2019){Drummond}, {Carter}, {H{\'e}brard},
  {Mayne}, {Sing}, {Evans}, \& {Goyal}}]{Drummond+19}
{Drummond}, B., {Carter}, A.~L., {H{\'e}brard}, E., {et~al.} 2019, \mnras, 486,
  1123, \dodoi{10.1093/mnras/stz909}

\bibitem[{{Edwards} {et~al.}(2022){Edwards}, {Changeat}, {Tsiaras}, {Hou Yip},
  {Al-Refaie}, {Anisman}, {Bieger}, {Gressier}, {Shibata}, {Skaf}, {Bouwman},
  {Y-K. Cho}, {Ikoma}, {Venot}, {Waldmann}, {Lagage}, \&
  {Tinetti}}]{Edwards+22}
{Edwards}, B., {Changeat}, Q., {Tsiaras}, A., {et~al.} 2022, arXiv e-prints,
  arXiv:2211.00649.
\newblock \doarXiv{2211.00649}

\bibitem[{{Eistrup}(2022)}]{Eistrup22}
{Eistrup}, C. 2022, arXiv e-prints, arXiv:2210.16921.
\newblock \doarXiv{2210.16921}

\bibitem[{{Eistrup} {et~al.}(2022){Eistrup}, {Cleeves}, \&
  {Krijt}}]{Eistrup+22}
{Eistrup}, C., {Cleeves}, L.~I., \& {Krijt}, S. 2022, arXiv e-prints,
  arXiv:2207.13158.
\newblock \doarXiv{2207.13158}

\bibitem[{{Eistrup} {et~al.}(2016){Eistrup}, {Walsh}, \& {van
  Dishoeck}}]{Eistrup+16}
{Eistrup}, C., {Walsh}, C., \& {van Dishoeck}, E.~F. 2016, \aap, 595, A83,
  \dodoi{10.1051/0004-6361/201628509}

\bibitem[{{Eistrup} {et~al.}(2018){Eistrup}, {Walsh}, \& {van
  Dishoeck}}]{Eistrup+18}
---. 2018, \aap, 613, A14, \dodoi{10.1051/0004-6361/201731302}

\bibitem[{{Espinoza} {et~al.}(2017){Espinoza}, {Fortney}, {Miguel},
  {Thorngren}, \& {Murray-Clay}}]{Espinoza+17}
{Espinoza}, N., {Fortney}, J.~J., {Miguel}, Y., {Thorngren}, D., \&
  {Murray-Clay}, R. 2017, \apjl, 838, L9, \dodoi{10.3847/2041-8213/aa65ca}

\bibitem[{{Fegley} \& {Prinn}(1985)}]{Fegley&Prinn85}
{Fegley}, B., J., \& {Prinn}, R.~G. 1985, \apj, 299, 1067,
  \dodoi{10.1086/163775}

\bibitem[{{Fegley} \& {Lodders}(1994)}]{Fegley&Lodders94}
{Fegley}, Bruce, J., \& {Lodders}, K. 1994, \icarus, 110, 117,
  \dodoi{10.1006/icar.1994.1111}

\bibitem[{{Fortney} {et~al.}(2011){Fortney}, {Ikoma}, {Nettelmann}, {Guillot},
  \& {Marley}}]{Fortney+11}
{Fortney}, J.~J., {Ikoma}, M., {Nettelmann}, N., {Guillot}, T., \& {Marley},
  M.~S. 2011, \apj, 729, 32, \dodoi{10.1088/0004-637X/729/1/32}

\bibitem[{{Fortney} {et~al.}(2008){Fortney}, {Lodders}, {Marley}, \&
  {Freedman}}]{Fortney+08}
{Fortney}, J.~J., {Lodders}, K., {Marley}, M.~S., \& {Freedman}, R.~S. 2008,
  \apj, 678, 1419, \dodoi{10.1086/528370}

\bibitem[{{Fortney} {et~al.}(2007){Fortney}, {Marley}, \&
  {Barnes}}]{Fortney+07}
{Fortney}, J.~J., {Marley}, M.~S., \& {Barnes}, J.~W. 2007, \apj, 659, 1661,
  \dodoi{10.1086/512120}

\bibitem[{{Fortney} {et~al.}(2005){Fortney}, {Marley}, {Lodders}, {Saumon}, \&
  {Freedman}}]{Fortney+05}
{Fortney}, J.~J., {Marley}, M.~S., {Lodders}, K., {Saumon}, D., \& {Freedman},
  R. 2005, \apjl, 627, L69, \dodoi{10.1086/431952}

\bibitem[{{Fortney} {et~al.}(2020){Fortney}, {Visscher}, {Marley}, {Hood},
  {Line}, {Thorngren}, {Freedman}, \& {Lupu}}]{Fortney+20}
{Fortney}, J.~J., {Visscher}, C., {Marley}, M.~S., {et~al.} 2020, \aj, 160,
  288, \dodoi{10.3847/1538-3881/abc5bd}

\bibitem[{{Freedman} {et~al.}(2014){Freedman}, {Lustig-Yaeger}, {Fortney},
  {Lupu}, {Marley}, \& {Lodders}}]{Freedman+14}
{Freedman}, R.~S., {Lustig-Yaeger}, J., {Fortney}, J.~J., {et~al.} 2014, \apjs,
  214, 25, \dodoi{10.1088/0067-0049/214/2/25}

\bibitem[{{Gao} {et~al.}(2020){Gao}, {Thorngren}, {Lee}, {Fortney}, {Morley},
  {Wakeford}, {Powell}, {Stevenson}, \& {Zhang}}]{Gao+20}
{Gao}, P., {Thorngren}, D.~P., {Lee}, G. K.~H., {et~al.} 2020, Nature
  Astronomy, \dodoi{10.1038/s41550-020-1114-3}

\bibitem[{{Ginzburg} {et~al.}(2018){Ginzburg}, {Schlichting}, \&
  {Sari}}]{Ginzburg+18}
{Ginzburg}, S., {Schlichting}, H.~E., \& {Sari}, R. 2018, \mnras, 476, 759,
  \dodoi{10.1093/mnras/sty290}

\bibitem[{{Guillot}(2010)}]{Guillot10}
{Guillot}, T. 2010, \aap, 520, A27, \dodoi{10.1051/0004-6361/200913396}

\bibitem[{{Guillot} {et~al.}(1996){Guillot}, {Burrows}, {Hubbard}, {Lunine}, \&
  {Saumon}}]{Guillot+96}
{Guillot}, T., {Burrows}, A., {Hubbard}, W.~B., {Lunine}, J.~I., \& {Saumon},
  D. 1996, \apjl, 459, L35, \dodoi{10.1086/309935}

\bibitem[{{Guillot} {et~al.}(2022){Guillot}, {Fletcher}, {Helled}, {Ikoma},
  {Line}, \& {Parmentier}}]{Guillot+22}
{Guillot}, T., {Fletcher}, L.~N., {Helled}, R., {et~al.} 2022, arXiv e-prints,
  arXiv:2205.04100.
\newblock \doarXiv{2205.04100}

\bibitem[{{Guillot} \& {Showman}(2002)}]{Guillot&Showman02}
{Guillot}, T., \& {Showman}, A.~P. 2002, \aap, 385, 156,
  \dodoi{10.1051/0004-6361:20011624}

\bibitem[{{Guillot} {et~al.}(2020{\natexlab{a}}){Guillot}, {Stevenson},
  {Atreya}, {Bolton}, \& {Becker}}]{Guillot+20}
{Guillot}, T., {Stevenson}, D.~J., {Atreya}, S.~K., {Bolton}, S.~J., \&
  {Becker}, H.~N. 2020{\natexlab{a}}, Journal of Geophysical Research
  (Planets), 125, e06403, \dodoi{10.1029/2020JE006403}

\bibitem[{{Guillot} {et~al.}(2020{\natexlab{b}}){Guillot}, {Li}, {Bolton},
  {Brown}, {Ingersoll}, {Janssen}, {Levin}, {Lunine}, {Orton}, {Steffes}, \&
  {Stevenson}}]{Guillot+20b}
{Guillot}, T., {Li}, C., {Bolton}, S.~J., {et~al.} 2020{\natexlab{b}}, Journal
  of Geophysical Research (Planets), 125, e06404, \dodoi{10.1029/2020JE006404}

\bibitem[{{Hands} \& {Helled}(2022)}]{Hands&Helled22}
{Hands}, T.~O., \& {Helled}, R. 2022, \mnras, 509, 894,
  \dodoi{10.1093/mnras/stab2967}

\bibitem[{{Hayashi} {et~al.}(1962){Hayashi}, {H{\={o}}shi}, \&
  {Sugimoto}}]{Hayashi+62}
{Hayashi}, C., {H{\={o}}shi}, R., \& {Sugimoto}, D. 1962, Progress of
  Theoretical Physics Supplement, 22, 1, \dodoi{10.1143/PTPS.22.1}

\bibitem[{{Helling} {et~al.}(2014){Helling}, {Woitke}, {Rimmer}, {Kamp}, {Thi},
  \& {Meijerink}}]{Helling+14}
{Helling}, C., {Woitke}, P., {Rimmer}, P.~B., {et~al.} 2014, Life, 4, 142,
  \dodoi{10.3390/life4020142}

\bibitem[{{Hobbs} {et~al.}(2019){Hobbs}, {Shorttle}, {Madhusudhan}, \&
  {Rimmer}}]{Hobbs+19}
{Hobbs}, R., {Shorttle}, O., {Madhusudhan}, N., \& {Rimmer}, P. 2019, \mnras,
  487, 2242, \dodoi{10.1093/mnras/stz1333}

\bibitem[{{Hu}(2019)}]{Hu19}
{Hu}, R. 2019, \apj, 887, 166, \dodoi{10.3847/1538-4357/ab58c7}

\bibitem[{{Hu}(2021)}]{Hu21}
---. 2021, \apj, 921, 27, \dodoi{10.3847/1538-4357/ac1789}

\bibitem[{JWST Transiting Exoplanet Community Early Release Science~Team
  {et~al.}(2022)JWST Transiting Exoplanet Community Early Release Science~Team,
  Alderson, Batalha, Batalha, Bean, Beatty, Bell, Benneke, Berta-Thompson,
  Carter, Crossfield, Espinoza, Feinstein, Fortney, Gibson, Goyal, Kempton,
  Kirk, Kreidberg, L{\'o}pez-Morales, Line, Lothringer, Moran, Mukherjee, Ohno,
  Parmentier, Piaulet, Rustamkulov, Schlawin, Sing, Stevenson, Wakeford, Allen,
  Birkmann, Brande, Crouzet, Cubillos, Damiano, D{\'e}sert, Gao, Harrington,
  Hu, Kendrew, Knutson, Lagage, Leconte, Lendl, MacDonald, May, Miguel,
  Molaverdikhani, Moses, Murray, Nehring, Nikolov, Petit dit de~la Roche,
  Radica, Roy, Stassun, Taylor, Waalkes, Wachiraphan, Welbanks, Wheatley,
  Aggarwal, Alam, Banerjee, Barstow, Blecic, Casewell, Changeat, Chubb,
  Col{\'o}n, Coulombe, Daylan, de~Val-Borro, Decin, Dos~Santos, Flagg, France,
  Fu, Mu{\~n}oz, Gizis, Glidden, Grant, Heng, Henning, Hong, Inglis, Iro,
  Kataria, Komacek, Krick, Lee, Lewis, Lillo-Box, Lustig-Yaeger, Mancini,
  Mandell, Mansfield, Marley, Mikal-Evans, Morello, Nixon, Ceballos, Piette,
  Powell, Rackham, Ramos-Rosado, Rauscher, Redfield, Rogers, Roman, Roudier,
  Scarsdale, Shkolnik, Southworth, Spake, Steinrueck, Tan, Teske, Tremblin,
  Tsai, Tucker, Turner, Valenti, Venot, Waldmann, Wallack, Zhang, \&
  Zieba}]{ERS+22}
JWST Transiting Exoplanet Community Early Release Science~Team, Ahrer, E.-M.,
  Alderson, L., Batalha, N.~M., {et~al.} 2022, Nature

\bibitem[{{Karalidi} {et~al.}(2021){Karalidi}, {Marley}, {Fortney}, {Morley},
  {Saumon}, {Lupu}, {Visscher}, \& {Freedman}}]{Karalidi+21}
{Karalidi}, T., {Marley}, M., {Fortney}, J.~J., {et~al.} 2021, \apj, 923, 269,
  \dodoi{10.3847/1538-4357/ac3140}

\bibitem[{{Kawashima} \& {Min}(2021)}]{Kawashima&Min21}
{Kawashima}, Y., \& {Min}, M. 2021, \aap, 656, A90,
  \dodoi{10.1051/0004-6361/202141548}

\bibitem[{{Kippenhahn} \& {Weigert}(1994)}]{Kippenhahn+94}
{Kippenhahn}, R., \& {Weigert}, A. 1994, {Stellar Structure and Evolution}

\bibitem[{{Kreidberg} {et~al.}(2014){Kreidberg}, {Bean}, {D{\'e}sert}, {Line},
  {Fortney}, {Madhusudhan}, {Stevenson}, {Showman}, {Charbonneau},
  {McCullough}, {Seager}, {Burrows}, {Henry}, {Williamson}, {Kataria}, \&
  {Homeier}}]{Kreidberg+14b}
{Kreidberg}, L., {Bean}, J.~L., {D{\'e}sert}, J.-M., {et~al.} 2014, \apjl, 793,
  L27, \dodoi{10.1088/2041-8205/793/2/L27}

\bibitem[{{Kubyshkina} {et~al.}(2020){Kubyshkina}, {Vidotto}, {Fossati}, \&
  {Farrell}}]{Kubyshkina+20}
{Kubyshkina}, D., {Vidotto}, A.~A., {Fossati}, L., \& {Farrell}, E. 2020,
  \mnras, 499, 77, \dodoi{10.1093/mnras/staa2815}

\bibitem[{{Kurokawa} \& {Nakamoto}(2014)}]{Kurokawa&Nakamoto14}
{Kurokawa}, H., \& {Nakamoto}, T. 2014, \apj, 783, 54,
  \dodoi{10.1088/0004-637X/783/1/54}

\bibitem[{{Kurosaki} {et~al.}(2014){Kurosaki}, {Ikoma}, \&
  {Hori}}]{Kurosaki+14}
{Kurosaki}, K., {Ikoma}, M., \& {Hori}, Y. 2014, \aap, 562, A80,
  \dodoi{10.1051/0004-6361/201322258}

\bibitem[{{Li} {et~al.}(2017){Li}, {Ingersoll}, {Janssen}, {Levin}, {Bolton},
  {Adumitroaie}, {Allison}, {Arballo}, {Bellotti}, {Brown}, {Ewald}, {Jewell},
  {Misra}, {Orton}, {Oyafuso}, {Steffes}, \& {Williamson}}]{Li+17}
{Li}, C., {Ingersoll}, A., {Janssen}, M., {et~al.} 2017, \grl, 44, 5317,
  \dodoi{10.1002/2017GL073159}

\bibitem[{{Line} {et~al.}(2011){Line}, {Vasisht}, {Chen}, {Angerhausen}, \&
  {Yung}}]{Line+11}
{Line}, M.~R., {Vasisht}, G., {Chen}, P., {Angerhausen}, D., \& {Yung}, Y.~L.
  2011, \apj, 738, 32, \dodoi{10.1088/0004-637X/738/1/32}

\bibitem[{{Lodders} \& {Fegley}(2002)}]{Lodders&Fegley02}
{Lodders}, K., \& {Fegley}, B. 2002, \icarus, 155, 393,
  \dodoi{10.1006/icar.2001.6740}

\bibitem[{{Lopez} \& {Fortney}(2014)}]{Lopez&Fortney14}
{Lopez}, E.~D., \& {Fortney}, J.~J. 2014, \apj, 792, 1,
  \dodoi{10.1088/0004-637X/792/1/1}

\bibitem[{{Lothringer} {et~al.}(2021){Lothringer}, {Rustamkulov}, {Sing},
  {Gibson}, {Wilson}, \& {Schlaufman}}]{Lothringer+21}
{Lothringer}, J.~D., {Rustamkulov}, Z., {Sing}, D.~K., {et~al.} 2021, \apj,
  914, 12, \dodoi{10.3847/1538-4357/abf8a9}

\bibitem[{Lupu {et~al.}(2021)Lupu, Freedman, \& Visscher}]{Lupu+21}
Lupu, R., Freedman, R., \& Visscher, C. 2021, {Correlated k coefficients for
  H2-He atmospheres; 196 spectral windows and 1060 pressure-temperature
  points},  Zenodo, \dodoi{10.5281/zenodo.6708165}

\bibitem[{{MacDonald} \& {Madhusudhan}(2017)}]{MacDonald&Madhusudhan17}
{MacDonald}, R.~J., \& {Madhusudhan}, N. 2017, \apjl, 850, L15,
  \dodoi{10.3847/2041-8213/aa97d4}

\bibitem[{{Madhusudhan}(2012)}]{Madhusudhan+12}
{Madhusudhan}, N. 2012, \apj, 758, 36, \dodoi{10.1088/0004-637X/758/1/36}

\bibitem[{{Madhusudhan} {et~al.}(2014){Madhusudhan}, {Amin}, \&
  {Kennedy}}]{Madhusudhan+14}
{Madhusudhan}, N., {Amin}, M.~A., \& {Kennedy}, G.~M. 2014, \apjl, 794, L12,
  \dodoi{10.1088/2041-8205/794/1/L12}

\bibitem[{{Madhusudhan} {et~al.}(2017){Madhusudhan}, {Bitsch}, {Johansen}, \&
  {Eriksson}}]{Madhusudhan+17}
{Madhusudhan}, N., {Bitsch}, B., {Johansen}, A., \& {Eriksson}, L. 2017,
  \mnras, 469, 4102, \dodoi{10.1093/mnras/stx1139}

\bibitem[{{Marley} \& {McKay}(1999)}]{Marley&McKay99}
{Marley}, M.~S., \& {McKay}, C.~P. 1999, \icarus, 138, 268,
  \dodoi{10.1006/icar.1998.6071}

\bibitem[{{Marley} \& {Robinson}(2015)}]{Marley&Robinson15}
{Marley}, M.~S., \& {Robinson}, T.~D. 2015, \araa, 53, 279,
  \dodoi{10.1146/annurev-astro-082214-122522}

\bibitem[{{Marley} {et~al.}(1996){Marley}, {Saumon}, {Guillot}, {Freedman},
  {Hubbard}, {Burrows}, \& {Lunine}}]{Marley+96}
{Marley}, M.~S., {Saumon}, D., {Guillot}, T., {et~al.} 1996, Science, 272,
  1919, \dodoi{10.1126/science.272.5270.1919}

\bibitem[{{Marley} {et~al.}(2021){Marley}, {Saumon}, {Visscher}, {Lupu},
  {Freedman}, {Morley}, {Fortney}, {Seay}, {Smith}, {Teal}, \&
  {Wang}}]{Marley+21}
{Marley}, M.~S., {Saumon}, D., {Visscher}, C., {et~al.} 2021, \apj, 920, 85,
  \dodoi{10.3847/1538-4357/ac141d}

\bibitem[{{Mayorga} {et~al.}(2021){Mayorga}, {Robinson}, {Marley}, {May}, \&
  {Stevenson}}]{Mayorga+21}
{Mayorga}, L.~C., {Robinson}, T.~D., {Marley}, M.~S., {May}, E.~M., \&
  {Stevenson}, K.~B. 2021, \apj, 915, 41, \dodoi{10.3847/1538-4357/abff50}

\bibitem[{{McKay} {et~al.}(1989){McKay}, {Pollack}, \& {Courtin}}]{McKay+89}
{McKay}, C.~P., {Pollack}, J.~B., \& {Courtin}, R. 1989, \icarus, 80, 23,
  \dodoi{10.1016/0019-1035(89)90160-7}

\bibitem[{{Miles} {et~al.}(2020){Miles}, {Skemer}, {Morley}, {Marley},
  {Fortney}, {Allers}, {Faherty}, {Geballe}, {Visscher}, {Schneider}, {Lupu},
  {Freedman}, \& {Bjoraker}}]{Miles+20}
{Miles}, B.~E., {Skemer}, A. J.~I., {Morley}, C.~V., {et~al.} 2020, \aj, 160,
  63, \dodoi{10.3847/1538-3881/ab9114}

\bibitem[{{Mizuno}(1980)}]{Mizuno80}
{Mizuno}, H. 1980, Progress of Theoretical Physics, 64, 544,
  \dodoi{10.1143/PTP.64.544}

\bibitem[{{Molaverdikhani} {et~al.}(2019){Molaverdikhani}, {Henning}, \&
  {Molli{\`e}re}}]{Molaverdikhani+19}
{Molaverdikhani}, K., {Henning}, T., \& {Molli{\`e}re}, P. 2019, \apj, 883,
  194, \dodoi{10.3847/1538-4357/ab3e30}

\bibitem[{{Molli{\`e}re} {et~al.}(2015){Molli{\`e}re}, {van Boekel},
  {Dullemond}, {Henning}, \& {Mordasini}}]{Molliere+15}
{Molli{\`e}re}, P., {van Boekel}, R., {Dullemond}, C., {Henning}, T., \&
  {Mordasini}, C. 2015, \apj, 813, 47, \dodoi{10.1088/0004-637X/813/1/47}

\bibitem[{{Molli{\`e}re} {et~al.}(2022){Molli{\`e}re}, {Molyarova}, {Bitsch},
  {Henning}, {Schneider}, {Kreidberg}, {Eistrup}, {Burn}, {Nasedkin},
  {Semenov}, {Mordasini}, {Schlecker}, {Schwarz}, {Lacour}, {Nowak}, \&
  {Schulik}}]{Molliere+22}
{Molli{\`e}re}, P., {Molyarova}, T., {Bitsch}, B., {et~al.} 2022, arXiv
  e-prints, arXiv:2204.13714.
\newblock \doarXiv{2204.13714}

\bibitem[{{Mordasini} {et~al.}(2012){Mordasini}, {Alibert}, {Klahr}, \&
  {Henning}}]{Mordasini+12}
{Mordasini}, C., {Alibert}, Y., {Klahr}, H., \& {Henning}, T. 2012, \aap, 547,
  A111, \dodoi{10.1051/0004-6361/201118457}

\bibitem[{{Morley} {et~al.}(2013){Morley}, {Fortney}, {Kempton}, {Marley},
  {Visscher}, \& {Zahnle}}]{Morley+13}
{Morley}, C.~V., {Fortney}, J.~J., {Kempton}, E.~M.-R., {et~al.} 2013, \apj,
  775, 33, \dodoi{10.1088/0004-637X/775/1/33}

\bibitem[{{Morley} {et~al.}(2012){Morley}, {Fortney}, {Marley}, {Visscher},
  {Saumon}, \& {Leggett}}]{Morley+12}
{Morley}, C.~V., {Fortney}, J.~J., {Marley}, M.~S., {et~al.} 2012, \apj, 756,
  172, \dodoi{10.1088/0004-637X/756/2/172}

\bibitem[{{Morley} {et~al.}(2015){Morley}, {Fortney}, {Marley}, {Zahnle},
  {Line}, {Kempton}, {Lewis}, \& {Cahoy}}]{Morley+15}
---. 2015, \apj, 815, 110, \dodoi{10.1088/0004-637X/815/2/110}

\bibitem[{{Morley} {et~al.}(2017){Morley}, {Knutson}, {Line}, {Fortney},
  {Thorngren}, {Marley}, {Teal}, \& {Lupu}}]{Morley+17}
{Morley}, C.~V., {Knutson}, H., {Line}, M., {et~al.} 2017, \aj, 153, 86,
  \dodoi{10.3847/1538-3881/153/2/86}

\bibitem[{{Morley} {et~al.}(2014){Morley}, {Marley}, {Fortney}, {Lupu},
  {Saumon}, {Greene}, \& {Lodders}}]{Morley+14}
{Morley}, C.~V., {Marley}, M.~S., {Fortney}, J.~J., {et~al.} 2014, \apj, 787,
  78, \dodoi{10.1088/0004-637X/787/1/78}

\bibitem[{{Moses} {et~al.}(2013{\natexlab{a}}){Moses}, {Madhusudhan},
  {Visscher}, \& {Freedman}}]{Moses+13}
{Moses}, J.~I., {Madhusudhan}, N., {Visscher}, C., \& {Freedman}, R.~S.
  2013{\natexlab{a}}, \apj, 763, 25, \dodoi{10.1088/0004-637X/763/1/25}

\bibitem[{{Moses} {et~al.}(2021){Moses}, {Tremblin}, {Venot}, \&
  {Miguel}}]{Moses+21}
{Moses}, J.~I., {Tremblin}, P., {Venot}, O., \& {Miguel}, Y. 2021, Experimental
  Astronomy, \dodoi{10.1007/s10686-021-09749-1}

\bibitem[{{Moses} {et~al.}(2011){Moses}, {Visscher}, {Fortney}, {Showman},
  {Lewis}, {Griffith}, {Klippenstein}, {Shabram}, {Friedson}, {Marley}, \&
  {Freedman}}]{Moses+11}
{Moses}, J.~I., {Visscher}, C., {Fortney}, J.~J., {et~al.} 2011, \apj, 737, 15,
  \dodoi{10.1088/0004-637X/737/1/15}

\bibitem[{{Moses} {et~al.}(2013{\natexlab{b}}){Moses}, {Line}, {Visscher},
  {Richardson}, {Nettelmann}, {Fortney}, {Barman}, {Stevenson}, \&
  {Madhusudhan}}]{Moses+13b}
{Moses}, J.~I., {Line}, M.~R., {Visscher}, C., {et~al.} 2013{\natexlab{b}},
  \apj, 777, 34, \dodoi{10.1088/0004-637X/777/1/34}

\bibitem[{{Mukherjee} {et~al.}(2022{\natexlab{a}}){Mukherjee}, {Batalha},
  {Fortney}, \& {Marley}}]{Mukherjee+22a}
{Mukherjee}, S., {Batalha}, N.~E., {Fortney}, J.~J., \& {Marley}, M.~S.
  2022{\natexlab{a}}, arXiv e-prints, arXiv:2208.07836.
\newblock \doarXiv{2208.07836}

\bibitem[{{Mukherjee} {et~al.}(2022{\natexlab{b}}){Mukherjee}, {Fortney},
  {Batalha}, {Karalidi}, {Marley}, {Visscher}, {Miles}, \&
  {Skemer}}]{Mukherjee+22b}
{Mukherjee}, S., {Fortney}, J.~J., {Batalha}, N.~E., {et~al.}
  2022{\natexlab{b}}, arXiv e-prints, arXiv:2208.14317.
\newblock \doarXiv{2208.14317}

\bibitem[{{Notsu} {et~al.}(2020){Notsu}, {Eistrup}, {Walsh}, \&
  {Nomura}}]{Notsu+20}
{Notsu}, S., {Eistrup}, C., {Walsh}, C., \& {Nomura}, H. 2020, \mnras, 499,
  2229, \dodoi{10.1093/mnras/staa2944}

\bibitem[{{Notsu} {et~al.}(2022){Notsu}, {Ohno}, {Ueda}, {Walsh}, {Eistrup}, \&
  {Nomura}}]{Notsu+22}
{Notsu}, S., {Ohno}, K., {Ueda}, T., {et~al.} 2022, \apj, 936, 188,
  \dodoi{10.3847/1538-4357/ac87fa}

\bibitem[{{{\"O}berg} \& {Bergin}(2016)}]{Oberg&Bergin16}
{{\"O}berg}, K.~I., \& {Bergin}, E.~A. 2016, \apjl, 831, L19,
  \dodoi{10.3847/2041-8205/831/2/L19}

\bibitem[{{{\"O}berg} \& {Bergin}(2021)}]{Oberg&Bergin21}
---. 2021, \physrep, 893, 1, \dodoi{10.1016/j.physrep.2020.09.004}

\bibitem[{{{\"O}berg} {et~al.}(2011){{\"O}berg}, {Murray-Clay}, \&
  {Bergin}}]{Oberg+11}
{{\"O}berg}, K.~I., {Murray-Clay}, R., \& {Bergin}, E.~A. 2011, \apjl, 743,
  L16, \dodoi{10.1088/2041-8205/743/1/L16}

\bibitem[{{{\"O}berg} \& {Wordsworth}(2019)}]{Oberg&Wordsworth19}
{{\"O}berg}, K.~I., \& {Wordsworth}, R. 2019, \aj, 158, 194,
  \dodoi{10.3847/1538-3881/ab46a8}

\bibitem[{{Ohno} \& {Fortney}(2022)}]{Ohno&Fortney22b}
{Ohno}, K., \& {Fortney}, J. 2022

\bibitem[{{Ohno} \& {Ueda}(2021)}]{Ohno&Ueda21}
{Ohno}, K., \& {Ueda}, T. 2021, \aap, 651, L2,
  \dodoi{10.1051/0004-6361/202141169}

\bibitem[{{Owen} \& {Wu}(2017)}]{Owen&Wu17}
{Owen}, J.~E., \& {Wu}, Y. 2017, \apj, 847, 29,
  \dodoi{10.3847/1538-4357/aa890a}

\bibitem[{{Owen} {et~al.}(1999){Owen}, {Mahaffy}, {Niemann}, {Atreya},
  {Donahue}, {Bar-Nun}, \& {de Pater}}]{Owen+99}
{Owen}, T., {Mahaffy}, P., {Niemann}, H.~B., {et~al.} 1999, \nat, 402, 269,
  \dodoi{10.1038/46232}

\bibitem[{{Pacetti} {et~al.}(2022){Pacetti}, {Turrini}, {Schisano}, {Molinari},
  {Fonte}, {Politi}, {Hennebelle}, {Klessen}, {Testi}, \&
  {Lebreuilly}}]{Pacetti+22}
{Pacetti}, E., {Turrini}, D., {Schisano}, E., {et~al.} 2022, arXiv e-prints,
  arXiv:2206.14685.
\newblock \doarXiv{2206.14685}

\bibitem[{{Piso} {et~al.}(2015){Piso}, {{\"O}berg}, {Birnstiel}, \&
  {Murray-Clay}}]{Piso+15}
{Piso}, A.-M.~A., {{\"O}berg}, K.~I., {Birnstiel}, T., \& {Murray-Clay}, R.~A.
  2015, \apj, 815, 109, \dodoi{10.1088/0004-637X/815/2/109}

\bibitem[{{Piso} {et~al.}(2016){Piso}, {Pegues}, \& {{\"O}berg}}]{Piso+16}
{Piso}, A.-M.~A., {Pegues}, J., \& {{\"O}berg}, K.~I. 2016, \apj, 833, 203,
  \dodoi{10.3847/1538-4357/833/2/203}

\bibitem[{{Polman} {et~al.}(2022){Polman}, {Waters}, {Min}, {Miguel}, \&
  {Khorshid}}]{Polman+22}
{Polman}, J., {Waters}, L.~B.~F.~M., {Min}, M., {Miguel}, Y., \& {Khorshid}, N.
  2022, arXiv e-prints, arXiv:2208.00469.
\newblock \doarXiv{2208.00469}

\bibitem[{{Prinn} \& {Barshay}(1977)}]{Prinn&Barshay77}
{Prinn}, R.~G., \& {Barshay}, S.~S. 1977, Science, 198, 1031,
  \dodoi{10.1126/science.198.4321.1031}

\bibitem[{{Ram{\'\i}rez} {et~al.}(2020){Ram{\'\i}rez}, {Cridland}, \&
  {Molli{\`e}re}}]{Ramirez+20}
{Ram{\'\i}rez}, V., {Cridland}, A.~J., \& {Molli{\`e}re}, P. 2020, \aap, 641,
  A87, \dodoi{10.1051/0004-6361/202038186}

\bibitem[{{Robinson} \& {Marley}(2014)}]{Robinson&Marley14}
{Robinson}, T.~D., \& {Marley}, M.~S. 2014, \apj, 785, 158,
  \dodoi{10.1088/0004-637X/785/2/158}

\bibitem[{{Rogers} \& {Seager}(2010)}]{Rogers&Seager10}
{Rogers}, L.~A., \& {Seager}, S. 2010, \apj, 712, 974,
  \dodoi{10.1088/0004-637X/712/2/974}

\bibitem[{{Rustamkulov} {et~al.}(2022){Rustamkulov}, {Sing}, {Mukherjee},
  {May}, {Kirk}, {Schlawin}, {Line}, {Piaulet}, {Carter}, {Batalha}, {Goyal},
  {L{\'o}pez-Morales}, {Lothringer}, {MacDonald}, {Moran}, {Stevenson},
  {Wakeford}, {Espinoza}, {Bean}, {Batalha}, {Benneke}, {Berta-Thompson},
  {Crossfield}, {Gao}, {Kreidberg}, {Powell}, {Cubillos}, {Gibson}, {Leconte},
  {Molaverdikhani}, {Nikolov}, {Parmentier}, {Roy}, {Taylor}, {Turner},
  {Wheatley}, {Aggarwal}, {Ahrer}, {Alam}, {Alderson}, {Allen}, {Banerjee},
  {Barat}, {Barrado}, {Barstow}, {Bell}, {Blecic}, {Brande}, {Casewell},
  {Changeat}, {Chubb}, {Crouzet}, {Daylan}, {Decin}, {D{\'e}sert},
  {Mikal-Evans}, {Feinstein}, {Flagg}, {Fortney}, {Harrington}, {Heng}, {Hong},
  {Hu}, {Iro}, {Kataria}, {Kempton}, {Krick}, {Lendl}, {Lillo-Box}, {Louca},
  {Lustig-Yaeger}, {Mancini}, {Mansfield}, {Mayne}, {Miguel}, {Morello},
  {Ohno}, {Palle}, {Petit dit de la Roche}, {Rackham}, {Radica},
  {Ramos-Rosado}, {Redfield}, {Rogers}, {Shkolnik}, {Southworth}, {Teske},
  {Tremblin}, {Tucker}, {Venot}, {Waalkes}, {Welbanks}, {Zhang}, \&
  {Zieba}}]{ERS+22_PRISM}
{Rustamkulov}, Z., {Sing}, D.~K., {Mukherjee}, S., {et~al.} 2022, arXiv
  e-prints, arXiv:2211.10487.
\newblock \doarXiv{2211.10487}

\bibitem[{{Saumon} {et~al.}(1995){Saumon}, {Chabrier}, \& {van
  Horn}}]{Saumon+95}
{Saumon}, D., {Chabrier}, G., \& {van Horn}, H.~M. 1995, \apjs, 99, 713,
  \dodoi{10.1086/192204}

\bibitem[{{Saumon} \& {Marley}(2008)}]{Saumon&Marley08}
{Saumon}, D., \& {Marley}, M.~S. 2008, \apj, 689, 1327, \dodoi{10.1086/592734}

\bibitem[{{Saumon} {et~al.}(2006){Saumon}, {Marley}, {Cushing}, {Leggett},
  {Roellig}, {Lodders}, \& {Freedman}}]{Saumon+06}
{Saumon}, D., {Marley}, M.~S., {Cushing}, M.~C., {et~al.} 2006, \apj, 647, 552,
  \dodoi{10.1086/505419}

\bibitem[{{Schneider} \& {Bitsch}(2021{\natexlab{a}})}]{Schneider&Bitsch21}
{Schneider}, A.~D., \& {Bitsch}, B. 2021{\natexlab{a}}, \aap, 654, A71,
  \dodoi{10.1051/0004-6361/202039640}

\bibitem[{{Schneider} \& {Bitsch}(2021{\natexlab{b}})}]{Schneider&Bitsch21b}
---. 2021{\natexlab{b}}, \aap, 654, A72, \dodoi{10.1051/0004-6361/202141096}

\bibitem[{{Stevenson}(1982)}]{Stevenson+82}
{Stevenson}, D.~J. 1982, \planss, 30, 755, \dodoi{10.1016/0032-0633(82)90108-8}

\bibitem[{{Tang} {et~al.}(2021){Tang}, {Robinson}, {Marley}, {Batalha}, {Lupu},
  \& {Prato}}]{Tang+21}
{Tang}, S.-Y., {Robinson}, T.~D., {Marley}, M.~S., {et~al.} 2021, \apj, 922,
  26, \dodoi{10.3847/1538-4357/ac1e90}

\bibitem[{{Thiabaud} {et~al.}(2015){Thiabaud}, {Marboeuf}, {Alibert}, {Leya},
  \& {Mezger}}]{Thiabaud+15}
{Thiabaud}, A., {Marboeuf}, U., {Alibert}, Y., {Leya}, I., \& {Mezger}, K.
  2015, \aap, 574, A138, \dodoi{10.1051/0004-6361/201424868}

\bibitem[{{Thorngren} \& {Fortney}(2019)}]{Thorngren&Fortney19}
{Thorngren}, D., \& {Fortney}, J.~J. 2019, \apjl, 874, L31,
  \dodoi{10.3847/2041-8213/ab1137}

\bibitem[{{Thorngren} {et~al.}(2019){Thorngren}, {Gao}, \&
  {Fortney}}]{Thorngren+19}
{Thorngren}, D., {Gao}, P., \& {Fortney}, J.~J. 2019, \apjl, 884, L6,
  \dodoi{10.3847/2041-8213/ab43d0}

\bibitem[{{Thorngren} {et~al.}(2016){Thorngren}, {Fortney}, {Murray-Clay}, \&
  {Lopez}}]{Thorngren+16}
{Thorngren}, D.~P., {Fortney}, J.~J., {Murray-Clay}, R.~A., \& {Lopez}, E.~D.
  2016, \apj, 831, 64, \dodoi{10.3847/0004-637X/831/1/64}

\bibitem[{{Toon} {et~al.}(1989){Toon}, {McKay}, {Ackerman}, \&
  {Santhanam}}]{Toon+89}
{Toon}, O.~B., {McKay}, C.~P., {Ackerman}, T.~P., \& {Santhanam}, K. 1989,
  \jgr, 94, 16287, \dodoi{10.1029/JD094iD13p16287}

\bibitem[{{Tsai} {et~al.}(2018){Tsai}, {Kitzmann}, {Lyons}, {Mendon{\c{c}}a},
  {Grimm}, \& {Heng}}]{Tsai+18}
{Tsai}, S.-M., {Kitzmann}, D., {Lyons}, J.~R., {et~al.} 2018, \apj, 862, 31,
  \dodoi{10.3847/1538-4357/aac834}

\bibitem[{{Tsai} {et~al.}(2017){Tsai}, {Lyons}, {Grosheintz}, {Rimmer},
  {Kitzmann}, \& {Heng}}]{Tsai+17}
{Tsai}, S.-M., {Lyons}, J.~R., {Grosheintz}, L., {et~al.} 2017, \apjs, 228, 20,
  \dodoi{10.3847/1538-4365/228/2/20}

\bibitem[{{Tsai} {et~al.}(2021){Tsai}, {Malik}, {Kitzmann}, {Lyons}, {Fateev},
  {Lee}, \& {Heng}}]{Tsai+21}
{Tsai}, S.-M., {Malik}, M., {Kitzmann}, D., {et~al.} 2021, \apj, 923, 264,
  \dodoi{10.3847/1538-4357/ac29bc}

\bibitem[{{Tsai} {et~al.}(2022){Tsai}, {Lee}, {Powell}, {Gao}, {Zhang},
  {Moses}, {H{\'e}brard}, {Venot}, {Parmentier}, {Jordan}, {Hu}, {Alam},
  {Alderson}, {Batalha}, {Bean}, {Benneke}, {Bierson}, {Brady}, {Carone},
  {Carter}, {Chubb}, {Inglis}, {Leconte}, {Lopez-Morales}, {Miguel},
  {Molaverdikhani}, {Rustamkulov}, {Sing}, {Stevenson}, {Wakeford}, {Yang},
  {Aggarwal}, {Baeyens}, {Barat}, {Borro}, {Daylan}, {Fortney}, {France},
  {Goyal}, {Grant}, {Kirk}, {Kreidberg}, {Louca}, {Moran}, {Mukherjee},
  {Nasedkin}, {Ohno}, {Rackham}, {Redfield}, {Taylor}, {Tremblin}, {Visscher},
  {Wallack}, {Welbanks}, {Youngblood}, {Ahrer}, {Batalha}, {Behr},
  {Berta-Thompson}, {Blecic}, {Casewell}, {Crossfield}, {Crouzet}, {Cubillos},
  {Decin}, {D{\'e}sert}, {Feinstein}, {Gibson}, {Harrington}, {Heng},
  {Henning}, {Kempton}, {Krick}, {Lagage}, {Lendl}, {Line}, {Lothringer},
  {Mansfield}, {Mayne}, {Mikal-Evans}, {Palle}, {Schlawin}, {Shorttle},
  {Wheatley}, \& {Yurchenko}}]{ERS+22_SO2}
{Tsai}, S.-M., {Lee}, E. K.~H., {Powell}, D., {et~al.} 2022, arXiv e-prints,
  arXiv:2211.10490.
\newblock \doarXiv{2211.10490}

\bibitem[{{Turrini} {et~al.}(2021){Turrini}, {Schisano}, {Fonte}, {Molinari},
  {Politi}, {Fedele}, {Pani{\'c}}, {Kama}, {Changeat}, \&
  {Tinetti}}]{Turrini+22}
{Turrini}, D., {Schisano}, E., {Fonte}, S., {et~al.} 2021, \apj, 909, 40,
  \dodoi{10.3847/1538-4357/abd6e5}

\bibitem[{{Valencia} {et~al.}(2013){Valencia}, {Guillot}, {Parmentier}, \&
  {Freedman}}]{Valencia+13}
{Valencia}, D., {Guillot}, T., {Parmentier}, V., \& {Freedman}, R.~S. 2013,
  \apj, 775, 10, \dodoi{10.1088/0004-637X/775/1/10}

\bibitem[{{Vazan} {et~al.}(2015){Vazan}, {Helled}, {Kovetz}, \&
  {Podolak}}]{Vazan+15}
{Vazan}, A., {Helled}, R., {Kovetz}, A., \& {Podolak}, M. 2015, \apj, 803, 32,
  \dodoi{10.1088/0004-637X/803/1/32}

\bibitem[{{Venot} {et~al.}(2013){Venot}, {H{\'e}brard}, {Ag{\'u}ndez},
  {Dobrijevic}, {Selsis}, {Hersant}, {Iro}, \& {Bounaceur}}]{Venot+13}
{Venot}, O., {H{\'e}brard}, E., {Ag{\'u}ndez}, M., {et~al.} 2013, in
  Astrophysics and Space Science Proceedings, Vol.~35, The Early Evolution of
  the Atmospheres of Terrestrial Planets, ed. J.~M. {Trigo-Rodriguez},
  F.~{Raulin}, C.~{Muller}, \& C.~{Nixon}, 67, \dodoi{K18-74538}

\bibitem[{{Visscher} {et~al.}(2006){Visscher}, {Lodders}, \&
  {Fegley}}]{Visscher+06}
{Visscher}, C., {Lodders}, K., \& {Fegley}, Bruce, J. 2006, \apj, 648, 1181,
  \dodoi{10.1086/506245}

\bibitem[{{Visscher} {et~al.}(2010){Visscher}, {Lodders}, \&
  {Fegley}}]{Visscher&Lodders10}
---. 2010, \apj, 716, 1060, \dodoi{10.1088/0004-637X/716/2/1060}

\bibitem[{{Wakeford} \& {Dalba}(2020)}]{Wakeford&Dalba20}
{Wakeford}, H.~R., \& {Dalba}, P.~A. 2020, Philosophical Transactions of the
  Royal Society of London Series A, 378, 20200054,
  \dodoi{10.1098/rsta.2020.0054}

\bibitem[{{Wakeford} {et~al.}(2017){Wakeford}, {Sing}, {Kataria}, {Deming},
  {Nikolov}, {Lopez}, {Tremblin}, {Amundsen}, {Lewis}, {Mandell}, {Fortney},
  {Knutson}, {Benneke}, \& {Evans}}]{Wakeford+17}
{Wakeford}, H.~R., {Sing}, D.~K., {Kataria}, T., {et~al.} 2017, Science, 356,
  628, \dodoi{10.1126/science.aah4668}

\bibitem[{{Welbanks} {et~al.}(2019){Welbanks}, {Madhusudhan}, {Allard},
  {Hubeny}, {Spiegelman}, \& {Leininger}}]{Welbanks&Madhusudhan19}
{Welbanks}, L., {Madhusudhan}, N., {Allard}, N.~F., {et~al.} 2019, \apjl, 887,
  L20, \dodoi{10.3847/2041-8213/ab5a89}

\bibitem[{{Zahnle} \& {Marley}(2014)}]{Zahnle+14}
{Zahnle}, K.~J., \& {Marley}, M.~S. 2014, \apj, 797, 41,
  \dodoi{10.1088/0004-637X/797/1/41}

\end{thebibliography}
%%%%%%%%%%%%%%%%%%%%%%%%%%%%%%%%%%%%%%%%%%%%%%%%%%
\end{document}